\documentclass[prd,tightenlines,nofootinbib,superscriptaddress]{revtex4}

\usepackage{amsfonts,amssymb,amsthm,bbm}
\usepackage{amsmath}

\usepackage{color,psfrag}
\usepackage[dvips]{graphicx}

\usepackage{subcaption}
\usepackage{graphicx}
\usepackage{xcolor}
\usepackage{tabularray}
\usepackage{physics}
\usepackage{csquotes}
\usepackage{booktabs}
\usepackage{float}

\usepackage{hyperref}  

\newcommand{\N}{{\mathbb N}}

\newcommand{\cN}{{\mathcal N}}

\newcommand{\cC}{{\mathcal C}}

\newcommand{\be}{\begin{equation}}
\newcommand{\ee}{\end{equation}}
\newcommand{\beq}{\begin{eqnarray}}
\newcommand{\eeq}{\end{eqnarray}}
\newcommand{\bes}{\begin{eqnarray}}
\newcommand{\ees}{\end{eqnarray}}

\newcommand{\lp}{\left(}
\newcommand{\rp}{ \right)}
\newcommand{\lc}{\left[}
\newcommand{\rc}{\right]}

\renewcommand{\sl}{{\mathfrak{sl}}}

\newcommand{\f}{\frac}

\def\nn{\nonumber}
\def\pp{\partial}

\def\rd{\mathrm{d}}

\def\vphi{\varphi}
\def\eps{\epsilon}
\def\om{\omega}

\def\hcC{\hat{\cC}}

\begin{document}

\title{Test-Field \textit{vs} Physical Quasi-Normal Modes in Scalar-Tensor Theories}


\author{{\bf Alexandre Arbey}}\email{alexandre.arbey@ens-lyon.fr}
\affiliation{Université Claude Bernard Lyon 1, CNRS/IN2P3, Institut de Physique des 2 Infinis de Lyon, UMR 5822, F-69622, Villeurbanne, France}

\author{\bf Etera R. Livine}\email{etera.livine@ens-lyon.fr}
\affiliation{ENS de Lyon, CNRS, Laboratoire de Physique (LPENSL), F-69342 Lyon, France}

\author{\bf Clara Montagnon}\email{clara.montagnon@ens-lyon.fr}
\affiliation{ENS de Lyon, CNRS, Laboratoire de Physique (LPENSL), F-69342 Lyon, France}

\begin{abstract}

\medskip

In the context of the general effort to model black hole dynamics, and in particular their return-to-equilibrium through quasi-normal modes, it is crucial to understand how much test-field perturbations deviate from physical perturbations in modified gravity scenarios.
On the one hand, physical perturbations follow the modified Einstein equations of the considered extension of general relativity. The complexity of those equations can quickly escalate with extra fields and non-linear couplings.
On the other hand, test-field perturbations, with negligible back-reaction on the space-time geometry, describe the propagation of both matter fields and spin $s=2$ gravitational waves on the black hole geometry. They are not subject to the intricacies of the modified Einstein equations, and only probe the background spacetime metric. If their physics were to not deviate significantly from physical perturbations, they would be especially useful to investigate predictions from quantum gravity scenarios which lack explicit detailed Einstein equations.

Here we focus on a specific  modified gravity solution -- BCL black holes in scalar-tensor theories --for which physical perturbations  and related QNM frequencies have already been studied and computed numerically. We compute the test-field QNM frequencies  and compare the two QNM spectra. This provides a concrete example of the significant differences arising between test-fields and physical  perturbations, and flags unphysical deviations related to the test-field framework.


\end{abstract}

\maketitle

{\small \tableofcontents }

\newpage

\section*{\uppercase{Introduction}}

So far General Relativity (GR) predictions are correctly matching with every experimental result, from the cosmological observations of the PLANCK collaboration \cite{Planck:2018nkj} to the dynamics of the Hulse-Taylor binary pulsar system \cite{Hulse:1974eb,Weisberg:1981bh}. The first direct detection of gravitational waves (GWs) emitted during the merger of two black holes (BHs) \cite{LIGOScientific:2016aoc} allows to test GR in the regime of strong gravity, as BHs are known for exhibiting the strongest gravitational fields. To this day, the GW data collected by the LIGO-Virgo-KAGRA collaboration is matching with the predictions of GR up to the measurement precision \cite{LIGOScientific:2016lio,LIGOScientific:2019fpa,LIGOScientific:2020tif}. The expected future new GW detectors -- such as LISA \cite{LISA:2017pwj,LISACosmologyWorkingGroup:2019mwx,LISA:2024hlh}, Einstein Telescope \cite{Punturo:2010zz,ET:2019dnz,Chiummo:2023kxt} and Cosmic Explorer \cite{Dwyer_2015,Evans:2021gyd} -- will allow for GW detections at a higher rate and with a better precision, leading to either the detection of deviations from GR or to its validation to a larger scale.

The waveform of a GW emitted during the merger of two BHs can be decomposed into three phases: the inspiral, the merger and the ringdown. Efforts are being made to model each of these regimes theoretically in order to compare those models with the GW measurements. In this paper we will focus on the ringdown phase, during which the GWs are emitted by the resulting BH as it settles down to a stationary state. The ringdown can thus be described by linear perturbations around a stationary BH and is therefore relatively easy to model. It turns out to be well described by a superposition of damped sinusoids, along with a GW burst right after the merge and a power-law tail at the end of the signal. The sum of damped sinusoids is characterised by complex frequencies, being referred to as the Quasi-Normal Modes (QNMs) \cite{Kokkotas:1999bd,Nollert:1999ji,Berti:2009kk,Konoplya:2011qq,franchini2023testing}. Their real parts represent the oscillating frequencies while their imaginary parts are related to the damping time. As the QNMs only depend on the BH parameters, their measurements, now referred to as \textit{Black Hole Spectroscopy}, is a great tool for the testing of GR and gives hope of finding quantum signatures \cite{Berti:2005ys,Berti:2018vdi,Maselli:2019mjd,LIGOScientific:2016lio,Ghosh:2021mrv}.

Despite the numerous experimental successes of GR, the theory is failing to provide a good description of the fate of BHs \cite{Penrose:1964wq} and of the very early universe \cite{Hawking:1970zqf}, amongst others. Efforts are then being made to build alternative theories of gravity. Two directions are usually considered: either extending GR or developing a mathematical formalism describing new microscopic structures. Most of the extensions of GR currently under scrutiny have an additional scalar degree of freedom compared to GR. They are referred as Scalar-Tensor theories \cite{Langlois:2018dxi,Kobayashi:2019hrl} and contain a number of BH solutions among other \cite{Babichev:2017guv,BenAchour:2018dap,Motohashi:2019sen,Charmousis:2019vnf,Minamitsuji:2019shy,BenAchour:2020wiw,Minamitsuji:2019tet,Anson:2020trg,BenAchour:2020fgy,Takahashi:2020hso,Babichev:2020qpr,Baake:2020tgk,Capuano:2023yyh,Bakopoulos:2023fmv,Babichev:2023psy}. 
On the other side, an example of a theory built using new microscopic space-time structures is Loop Quantum Gravity (LQG). It is a non-perturbative and background independent approach to quantizing GR, and it is now a promising candidate for a theory of Quantum Gravity. It is nevertheless still very challenging to extract predictions directly from the full LQG theory. The use of effective BH models, build using concepts, tools and methods from LQG and expressed as quantum corrected Schwarzschild solutions, has then become common practice. 
More generally, numerous effective models of BH solutions have been derived, with different aims and different techniques. We can mention, for example, the study of BH effective models as dark matter \cite{Calza:2024fzo,Calza:2024xdh}.

The theoretical study of QNMs for effective BH solutions is of great interest nowadays, as it helps learning about the potential quantum corrections that could be detected via the BH spectroscopy. It is also a way to test the different effective BH metrics and discriminate them. However, those effective solutions do not come with modified Einstein equations and the usual process of metric perturbations cannot be performed for those BHs. It is, therefore, a common practice to either study scalar (spin $s=0$) test-field QNMs -- determined thanks to the Klein-Gordon equation and depending only on the BH metric -- or gravitational (spin $s=2$) test-field QNMs computed using the GR Einstein equation. The question is then: how good approximation are the spin $s=2$ test-field perturbations for the description of the proper physical gravitational perturbations? In other words, we want to question the relative importance of BH metrics corrections \textit{versus} space-time dynamics corrections. 
A first step towards the answer is the QNM study of one BH solution from Scalar-Tensor theories for which gravitational perturbations have already been studied: the BCL BH \cite{Babichev:2017guv}. The perturbation equations for the BCL BH have been derived in \cite{Langlois:2021xzq} and its axial gravitational QNMs have been numerically computed in \cite{Roussille:2023sdr}. Our aim here is then to look at the test-field QNMs and to see to what extent they differ from the physical ones. 

Various semi-analytical and numerical methods have been developed for QNM computation. One of the most efficient and accurate techniques has been proposed by Leaver in 1985 \cite{Leaver:1985ax} and relies on a reformulation of the eigenvalue problem into a question of finding the roots of a continued fraction.
A key point of the method is to impose the appropriate physically motivated boundary conditions: the modes are ingoing at the BH horizon and outgoing at spatial infinity.
Leaver's approach, often referred to as the \enquote{Continued Fraction method}, allows for a high-accuracy computation of QNMs on a large part of the complex spectrum \cite{Onozawa:1996ux, Berti:2003jh, Berti:2004um, Leaver:1990zz, Berti:2005eb, Yoshida:2003zz,Moreira:2023cxy,Livine:2024bvo}. This numerical method will be our first tool for the study of test-field QNMs of the BCL BH. We will associate it with an analytical study of the highly damped QNMs via the monodromy technique \cite{Motl:2002hd,Motl:2003cd,Natario:2004jd,Dreyer:2002vy,Konoplya:2011qq} and the near-horizon $\sl_{2}$ symmetry \cite{Bertini:2011ga}. We also use another numerical method, the WKB one \cite{Schutz:1985km, Iyer:1986np, Iyer:1986nq, Konoplya:2003ii,Matyjasek:2017psv,Konoplya:2019hlu}, as a cross-check for the low damped QNMs. 

\medskip

This paper is structured as follows. In the first section, we present the BCL BH solution in Scalar-Tensor theories. In the second section, we review the formalism used for the derivation of the Teukolsky equations for test-fields of spin 0 and 2. We then describe the continued fraction method, providing a powerful way to numerically compute QNM frequencies on a large part of the complex spectrum. In section III, we explore analytical techniques for inferring the asymptotic part of the QNM spectrum. While this acts as a cross-check of the numerical computations, this  also allow us to gain a better understanding of the symmetries and features of the BCL black hole and its perturbations. Then in the section IV, we give an overview of our numerical results for spin 0 and 2 test-field QNMs and compare them with the analytical computations.
We  provide an cross-check with the WKB approximation detailed in appendix.
Finally, in section V, we review the steps leading to the equation describing the physical axial perturbations and we compare the effective potential arising from this equation to the one obtained in the test-field equation. Finally, we compare  directly the  QNM spectra and analyze the extent of the deviation between physical and test fields, as we vary the BCL parameter $r_-$ and the number of modes.
We conclude with perspectives and outlook.

\section{BCL BH solution in Scalar-Tensor Theories}


In this work we focus on Scalar-Tensor theories \cite{Langlois:2018dxi,Kobayashi:2019hrl}, which provide a natural and broad class of extensions of General Relativity. Indeed, most theories of modified gravity admit a reformulation as a Scalar-Tensor theory in some regime. These theories are characterized by an additional scalar degree of freedom (which can be either fundamental or effective) compared to General Relativity and are often intended for modelling dark energy and the early Universe.

The most general Scalar-Tensor theory with second order equations of motion was developed by Horndeski in the 1970's \cite{Horndeski:1974wa}.
The interested reader will find a comprehensive modern review of degenerate higher order scalar-tensor (DHOST) theories in \cite{BenAchour:2024hbg}.
Here we consider a subset of Horndeski theories for which the action does not contain any cubic term in the second derivative of the scalar field $\phi$ and which is invariant under a constant shift of the scalar field. Those Scalar-Tensor theories are referred as the quadratic shift-symmetric Horndeski theories and are described by the action
\be
S\lc g_{\mu\nu},\phi\rc=\int \dd^4x \sqrt{-g}\lc F(X)R+P(X)+Q(X)\Box\phi + 2\frac{\partial F}{\partial X}\lp \phi_{\mu\nu}\phi^{\mu\nu}-(\Box\phi)^2\rp\rc,
\label{BCL_action}
\ee
where $X=\nabla^{\mu}\phi\nabla_{\mu}\phi$ and $\phi_{\mu\nu}=\nabla_{\mu}\nabla_{\nu}\phi$.
Here we will restrict to a special case where 
\be
F(X)=F_0+F_1\sqrt{X}, \ \ \ P(X)=-P_1X, \ \ \ Q(X)=0,
\label{action_fct}
\ee
with $F_0,F_1,P_1$ being constant. 
\\
This restriction is motivated by the existence for this case of a BH solution, refereed to as the BCL solution, from the name of its authors \cite{Babichev:2017guv}. This BH solution stands out by the simplicity of its metric functions -- resembling a lot the Reissner-Nordström metric functions -- while the deformations from the Schwarzschild case are wrapped into a single parameter (being $r_-$ or equivalently $\xi$, as we shall see). Moreover, the BCL solution is one of the few BH solutions from modified gravity for which the physical gravitational perturbations have been computed \cite{Langlois:2021xzq} and it is, up to this day, the only one for which a large part of the axial QNM spectrum has been numerically computed \cite{Roussille:2023sdr}. This gives a unique chance to compare the QNM spectra of a BH solution from modified gravity for, on the one side, physical gravitational perturbations and, on the other side, test-field perturbations.\\
The BCL solution for Scalar-Tensor is described by the metric \cite{Langlois:2021xzq}
\be
\dd s^2=-f(r)\dd t^2+\frac{1}{f(r)}\dd r^2+h(r)\dd \Omega^2,
\ee
with 
\be
f(r)=\lp 1-\frac{r_+}{r}\rp\lp 1+\frac{r_-}{r}\rp, \ \ \ h(r)=r^2.
\label{metric_func}
\ee
One can notice that this metric does not solve the singularity at $r=0$.\\
The quantities $r_+$ and $r_-$ are related to the functions \eqref{action_fct} and to the Schwarzschild radius by
\be 
r_+r_- = \frac{F_1^2}{2F_0P_1}, \ \ \ \ r_+-r_-=r_s, \ \ \ \ r+ > r_- > 0.
\ee 
The metric function $f(r)$ can also be written as
\be 
f(r)=1-\frac{r_s}{r}-\frac{\xi r_s^2}{2r^2}, \ \ \ \text{with \ } \xi=2\frac{r_+r_-}{r_s^2}=\frac{F_1^2}{F_0P_1r_s^2}.
\ee
The shape of the BCL metric resembles the Reissner-Nordström metric from General Relativity. The major difference is that Reissner-Nordström BH has two physical horizons, while the BCL BH only has one horizon of radius $r_+$, being the event horizon of the BCL BH. The parameter $r_-$ appearing here does not have any physical meaning in terms of horizon or special length. It is simply a deformation parameter, as $\xi$ is, but we will keep the formulation with $r_-$ for simplicity and similarity with the Reissner-Nordström metric. 
\\
The scalar field is given by
\be 
\phi(r)=\pm \frac{F_1}{P_1\sqrt{r_+r_-}}\arctan \lc \frac{r_s r+2r_+r_-}{2\sqrt{r_+r_-}\sqrt{(r-r_+)(r+r_-)}}\rc + \text{cst}.
\ee 
Note that the sign and the constant shift are not physically relevant as the action only depends on squared derivatives of $\phi$. This constant is a pure gauge.
\\
One can recover the Schwarzschild solution by taking $r_-$ (or equivalently $\xi$) to zero. Indeed, those quantities are playing the role of a single deformation parameter from the GR static and spherically symmetric BH solution. In the following, we will study the perturbations of the BCL BH as a mathematical toy model without intentions of physical predictions, and we will use the parameter $r_-$ as a marker of deviations from GR. 

\section{Test-field perturbations for massless spin 0 and 2}

\subsection{Teukolsky equations for massless spin 0 and 2}

In this work, we are interested in QNMs for a massless scalar field, with spin 0, and for a massless spin 2 field. Indeed, on the one hand, the scalar field is mathematically the simplest type of matter and is the proof-of-concept case for all type of fields. On the other hand, the spin-2 field is the test field approximation for physical gravitational perturbations and is the main focus of our analysis aiming at understanding the differences between physical perturbations and test-field perturbations for black holes in modified gravity theories.
We follow the logic and calculations of \cite{Arbey:2021jif}, extending many earlier works, in deriving the Teukolsky equations for the propagation of test fields of arbitrary spins in spherically-symmetric metric background.

Starting with spin 0, a classical massless scalar field $\phi$ follows the Klein-Gordon equation:
\be
\f1{\sqrt{-g}}\pp_{a}
\big{(}g^{{ab}}\sqrt{-g}\pp_{b}\phi\big{)}
=0\,.
\ee
Assuming that we study a test-field, that is without feedback on the metric, we plug directly the black hole metric in this equation. In particular $\sqrt{-g}=h\sin\theta$ does not depend on the metric component $f(r)$. Decomposing the scalar field in spherical harmonics $S^{0}_{l,m}$,
\be
\phi=\Phi_{0}(r)S^{0}_{l,m}(\theta,\vphi)e^{-i\om t}\,,
\ee
the Proca equation leads to a radial Teukolsky equation:
\be
\lp
\om^{2}-\f{l(l+1)f}{h}
\rp
\Phi_{0}
+\f{f}{h}\pp_{r}\lp
fh\pp_{r}\Phi_{0}
\rp
=0
\,.
\ee
Using explicitly that $h(r)=r^{2}$, one can  rescale the scalar field in order to absorb the $f h$ factor between the two radial derivatives. This allows to write the equation for $\Psi_{0}\equiv r \Phi_{0}$ as a second order differential equation in the tortoise coordinate $x$ defined by $\pp_{x}=f(r)\pp_{r}$ or equivalently $\rd x=f(r)^{-1}\rd r$:
\be
\pp_{x}^{2}\Psi_{0}+\Big{[}\om^{2}-V_{0}(r)\Big{]}\,\Psi_{0}=0\,,
\ee
with the effective potential:
\be
V_{0}
=
l(l+1)\f f h +\f1{\sqrt{h }}\pp^{2}_{x}\sqrt{h}
=
\f {f(r)}r
\lp
\f{l(l+1)}r +\f{r_{+}-r_{-}}{r^{2}}+2\f{r_{-}r_{+}}{r^{3}}
\rp
\,.
\ee

\medskip

Now turning to the spin 2 test-field, the point is to translate the field equations coupling all the components of the spin 2 field into decoupled equations, thus starting the spin 2 field as a collection of scalar fields propagating in the same metric background but with a modified effective potential. The recent work \cite{Arbey:2021jif} follows the original prescription by Teukolsky introduced in \cite{Teukolsky:1973ha} based on the Newman-Penrose formalism. We introduce a null tetrad adapted to the black hole metric:
\be
l^{a}=\lp f^{-1},1,0,0 \rp
\,,\quad
n^{a}=\f12\lp 1,-f,0,0 \rp
\,,\quad
m^{a}=\f1{2h}\lp 0,0,1,\f i{\sin\theta} \rp
\,,\quad
\bar{m}^{a}=\f1{2h}\lp 0,0,1,\f {-i}{\sin\theta} \rp
\,,
\ee
satisfying $g^{ab}=-l^{a}n^{b}-n^{a}l^{b}+m^{a}\bar{m}^{b}+\bar{m}^{a}m^{b}$ and the following scalar products:
\be
l\cdot n=-1, \ \ m\cdot \bar{m}=1, \ \ 
l\cdot l=n\cdot n=m\cdot m=\bar{m}\cdot \bar{m}=0, \ \ 
l\cdot m=l\cdot \bar{m}=n\cdot m=n\cdot \bar{m}=0.
\ee
Assuming that one studies metric perturbations, the field equations for those gravitational perturbations inheriting the standard Einstein equations will give the field equations for a spin-2 test-field. If one were to consider, instead, the modified Einstein equations of the considered modified gravity theory, we would get field equations for physical gravitational perturbations.

Thus, considering the standard Newman-Penrose equations, giving the projection of the standard Einstein equations on the null tetrad, one obtains equations for the perturbed components of the Weyl tensor $C_{abcd}$. More particularly, one gets a decoupled field equation for the component $\phi_{2}=-C_{abcd}l^{a}m^{b}l^{c}m^{d}$. The other components can then obtained as derivatives of that primary field.
Decomposing  the field $\phi_{2}$ on spin-2 weighted spherical harmonics $S^{2}_{l,m}(\theta,\vphi)$,
\be
\phi_{2}=\Phi_{2}(r)S^{2}_{l,m}(\theta,\vphi)e^{-i\om t}\,,
\ee
one obtains the radial Teukolsky equation for $\Phi_{2}(r)$. Finally, rescaling that radial component as $\Psi_{2}=r^{3}f\Phi_{2}$, one write it in a similar fashion as for the scalar field:
\be
\pp_{x}^{2}\Psi_{2}+\Big{[}\om^{2}-V_{2}(r)\Big{]}\,\Psi_{2}=0\,,
\ee
with the effective potential:
\be
V_{2}
=
\big{[}l(l+1)-2\big{]}\f fh
+\f12\left(\f{\pp_{x}h}{h}\right)^{2}
-\f1{\sqrt{h}}\pp^{2}_{x}\sqrt{h}
=
\f {f(r)}r
\lp
\f{l(l+1)}r -3\f{(r_{+}-r_{-})}{r^{2}}-4\f{r_{-}r_{+}}{r^{3}}
\rp
\,.
\ee
One sees that this potential driving the evolution of spin-2 fields has the same shape as the one driving the evolution of scalar fields, but has clearly distinct numerical factors. Thus we expect similar qualitative behaviour, with different numerics.

\subsection{Computation of QNMs using Leaver's continued fractions}
\label{Leaver_section}

\begin{figure}[!ht]
    \centering
    \begin{subfigure}{0.49\textwidth}
        \centering
        \includegraphics[width=0.85\linewidth]{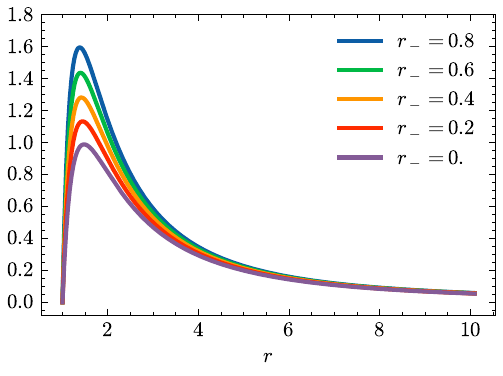}
        \caption{$s=0$.}
    \end{subfigure}
    \begin{subfigure}{0.49\textwidth}
        \centering
        \includegraphics[width=0.85\linewidth]{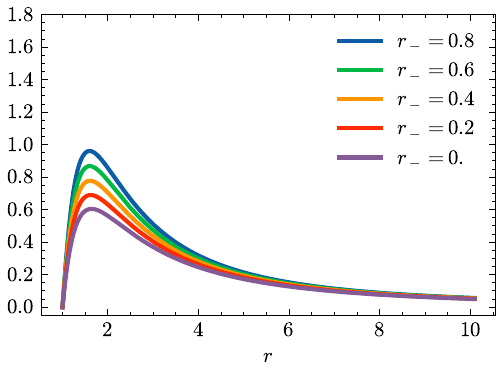}
        \caption{$s=2$.}
    \end{subfigure}
    \caption{Evolution of the BCL effective potential for spin $s=0$ and $s=2$, with the deformation parameter $r_-$ spanning from 0 to 0.8. The $r_-=0$ case correspond to the Schwarzschild BH}
    \label{BCL_pot_plots}
\end{figure}

We want to look at spin $s=0$ and $s=2$ test-field perturbations in the background of the BCL BH solution described above. The results we will display from now on are valid for $s=0$ and $s=2$, the only two cases we are interested in for this paper. The generalization to other values of the spin is left to the reader.\\
As we saw in the previous section, the dynamic of a massless test-field $\Psi_s$ of spin $s$ is given by the equation:
\be 
\frac{\dd^2 \Psi_s}{\dd x^2} + [\omega^2-V_s(r)]\Psi_s = 0,
\label{schro}
\ee
where the effective potential $V_s$ is given by
\be 
V_s(r)=\frac{f(r)}{r}\lp\frac{\lambda}{r}+(1-s^2)\frac{r_+-r_-}{r^2}+(2-3s)\frac{r_-r_+}{r^3}\rp.
\label{BCL_pot}
\ee 
\ \\
The angular momentum contribution is written as $\lambda=l(l+1)$. One recovers the Schwarzschild effective potential (\cite{Leaver:1985ax,Livine:2024bvo}) in the limit $r_-=0$.\\
From now on, let us set $r_+$ to 1 for convenience. The deviation parameter $r_-$ is now strictly bounded between 0 and 1: $0<r_-<1$. The evolution of the effective potential $V_s$ \eqref{BCL_pot} with respect to the radial coordinate $r$ for different values of the parameter $r_-$ and for spin $s=0$ and 2 is shown on figure \ref{BCL_pot_plots}. One can see that in both case, the height of the maxima of the potential is raised with $r_-$. Moreover, one can notice that the effective potential for spin 0 is more sensitive to shifts of $r_-$ than the one for spin 2. We expect this difference of sensitivity to manifest in the QNM spectra.
\\

As in the Schwarzschild case, the effective potential $V_s$ \eqref{BCL_pot} vanishes both at the horizon $r = r_+$  and at radial infinity $r \to +\infty$. Given the shape of the equation describing the test-fields perturbations, we can deduce the existence of asymptotic plane-waves in the tortoise coordinate $x$.
Note that the tortoise coordinate is defined as earlier by the differential equation $\dd x=f^{-1}(r) \ \dd r$, in terms of the metric function $f(r)$ given by \eqref{metric_func}. Integrating this equation gives
\be 
x(r) = r+\frac{r_+^2\ln(r-r_+)-r_-^2\ln(r+r_-)}{r_++r_-} + \text{cste}.
\ee
The asymptotic behaviour of this tortoise coordinate $x$ is then such that
\begin{subequations}
    \begin{align}
        &x = r + (r_+ - r_-) \ln(r) + \mathcal{O}(1) \,, &\mkern-120mu(r \to +\infty)\\
        &x = \frac{r_+^2}{r_++r_-} \ln(r - r_+) + \mathcal{O}(1)  \,. &\mkern-120mu(r \to r_+)
    \end{align}
    \label{eq:asymp-tortoise}
\end{subequations}

These asymptotic expressions for the tortoise coordinate $x$ can then be used to express the plane-wave asymptotic behaviour of the wave-function $\Psi_s$.
Moreover, QNMs are defined such that the wave is only ingoing at the BH event horizon and only outgoing at infinity. Imposing those boundary conditions leads to the following asymptotic behaviour:
\begin{subequations}
    \begin{align}
        &\Psi_s \sim e^{i\omega x} \sim e^{i\omega r}r^{i\omega(r_+ -r_-)} \,, &\mkern-120mu(r \to +\infty)\\
        &\Psi_s \sim e^{-i\omega x} \sim (r-r_+)^{-i\omega\frac{r_+^2}{r_++r_-}}  \,. &\mkern-120mu(r \to r_+)
        \label{boundaryhorizon}
    \end{align}
    \label{asympto}
\end{subequations}

Let us now turn to the computation of test-fields QNMs for the BCL BH using the Continued Fraction method developed by Leaver in 1985 \cite{Leaver:1985ax} and reviewed in details in our previous paper \cite{Livine:2024bvo}. The first step is to build an ansatz for the field function $\Psi_s$ as a power series, satisfying both the asymptotic behaviours at the horizon and at infinity \eqref{asympto}. The QNM domain of definition being $(r_+,+\infty)$, a power series of $r$ is not appropriate. The mapping
\be 
r \mapsto \frac{r-r_+}{r+r_-}
\ee
allows to shift the domain to $(0,1)$ and to obtain a well-defined power-series. We thus expand the field as:
\begin{equation}
\label{ansatz}
\Psi_s(r) = e^{i\omega(r-r_+)}r^{i\omega(r_+-r_-)}
\lp \frac{r-r_+}{r+r_-}\rp^{-i\omega\frac{r_+^2}{r_++r_-}}
\sum_{n=0}^{\infty}a_n\lp \frac{r-r_+}{r+r_-}\rp^n
\,,
\end{equation}
which reduces to the Schwarzschild ansatz (see \cite{Leaver:1985ax, Livine:2024bvo}) when  $r_-$ is taken to 0.

This expansion automatically implements the desired asymptotic behaviour at the near horizon and at large radial coordinate.
Let us underline that this is actually true only if the power series is well-defined. This means that the QNM boundary conditions directly translate into conditions on the coefficient sequence $\{a_n\}_{n\in\N}$:
\begin{itemize}
\item { QNM boundary condition at \textbf{horizon}}:
we require that $a_0\ne 0$ and that there is no negative power terms, $a_{n}=0$ for $n<0$.

\item {QNM boundary condition at \textbf{infinity}}:
As $(r-r_+)/(r+r_-)\sim 1$ as $r\rightarrow+\infty$,
we require that the series $\sum a_n$ is convergent.

\end{itemize}

%
Plugging the ansatz \eqref{ansatz} for the test-field perturbations of the BCL BH \eqref{ansatz} in the second-order differential equation \eqref{schro} yields recursion equations that must be satisfied by the coefficients $a_n$: 
%
\begin{equation}
\label{recur_eq}
\left|
    \begin{array}{rcl}
        \alpha_0a_1+\beta_0a_0&=&0
        \,,\\
        \alpha_1a_2+\beta_1a_1+\gamma_1a_0&=&0
        \,,\\
        \alpha_2a_3+\beta_2a_2+\gamma_2a_1+\delta_2a_0&=&0
        \,,\\
        \alpha_n a_{n+1}+\beta_na_n+\gamma_na_{n-1}+\delta_na_{n-2}+\epsilon_na_{n-3}
        &=&0\,, \qquad\textrm{for}\quad n\geq 3.
    \end{array}
\right.
\end{equation}
The coefficients $\alpha_n, \beta_n, \gamma_n, \delta_n, \epsilon_n$ are all quadratic polynomials in $n$. They also depend on the BCL BH deformation parameter $r_-$, the frequency $\omega$, the spin $s$, and the angular momentum parameter $\lambda$.
%
They read explicitly, for spins $s=0$ and $s=2$ (and $r_+=1$):
\beq
\alpha_n &=& (r_-+1)(n+1)^2 - 2i\omega (n+1)
\,,
\\
\beta_{n}&=&
2(r_--1)(r_-+1)n^2 + \lc-2i\omega r_-^3 -2r_-^2 + 2i(2i+\omega)r_- +8i\omega -2\rc n -i\omega r_-^3 + (-2\omega^2 + s-1)r_-^2
\nn\\
&&+ (2\omega^2+5i\omega - \lambda +3s-2)r_- + 8\omega^2+4i\omega -\lambda + s^2-1
\,,
\nn\\
\gamma_{n} &=&
(r_--2-\sqrt{3})(r_--2+\sqrt{3})(r_-+1)n^2 - 2i\lc\omega r_-^4 - 2(\omega+i)r_-^3+(-5\omega+2i)r_-^2 + (-6\omega+4i)r_- +2\omega\rc n
\nn\\
&&-\omega^2 r_-^5 +\omega(\omega+4i)r_-^4 + (3\omega^2 -2s+4)r_-^3 + (5\omega^2 - 8i\omega -2\lambda -4s +2)r_-^2
\nn\\
&&-2(-6\omega^2+8i\omega+\lambda+s^2+1)r_- -4\omega^2 -s^2
\,,
\nn\\
\delta_n &=&
-2r_-(r_--1)(r_-+1)n^2 + 2r_-\left[ 2i\omega r_-^3 + (3i\omega+5)r_-^2+(2i\omega+2)r_- -3i\omega -3\rc n
\nn\\
&&
+r_-\lc 2\omega^2 r_-^4 + 2\omega(2\omega-5i)r_-^3 - (15i\omega + \lambda -2s +13)r_-^2 \right.
\nn\\
&&\left. -(-6\omega^2 +10i\omega+\lambda -3s+10)r_- - 4\omega^2+11i\omega+s+3\right]
\,,
\nn\\
\eps_n &=&
 r_-^2(r_-+1)n^2 - 2(3r_-+i\omega+3)r_-^2 n + r_-^2\lc \omega^2(r_-^3-r_-^2+r_--1)+9(r_-+1)+6i\omega \rc\,.
 \nn
\eeq

\medskip

Let us first describe the Schwarzschild case, when $r_-$ vanishes.
The coefficients $\delta_n$ and $\epsilon_n$ identically vanish and one falls back to the standard 3-term recursion relation of Schwarzschild BHs, already well-studied (see e.g. \cite{Leaver:1985ax, Livine:2024bvo}):
\be
\alpha_n a_{n+1}+\beta_na_n+\gamma_na_{n-1}=0\,,
\ee
with the recursion coefficients, conveniently given in terms of $\rho\equiv -i\om$,
\beq
\alpha_n &=& n^{2}+(2\rho+2)n+2\rho+1
\,,
\\
\beta_{n}&=& -\big{[}
2n^{2}+(8\rho+2)n+8\rho^{2}+4\rho+\lambda-s^{2}+1
\big{]}
\,,
\nn\\
\gamma_{n} &=& n^{2}+4\rho^{2}+4\rho-s^{2}
\,.
\nn
\eeq
In that setting, Leaver's method consists in recasting the conditions on the power series coefficients $\{a_n\}_{n\in\N}$ in terms of the ratios $R_n\equiv a_n/a_{n-1}$.
First, as we have explained earlier, the asymptotic QNM boundary condition at $r\rightarrow+\infty$ amounts to requiring the convergence of the series $\sum a_n$. As pointed out in Leaver's original paper, the ratios $R_n$ converge to 1 for all values of the frequency $\om$. This is easily confirmed by an inverse power expansion in $n$ of the recursion relation, yielding:
\be
\alpha_n R_{n+1}+\beta_n+\gamma_n \f1{R_{n}}=0
\,\,\Rightarrow\,\,
R_{n}
\underset{n\gg1}\sim1\pm\sqrt{\f{2\rho}{n}}+\f{2\rho-\f34}{n}+\frac{3+16\lambda-48\rho+64\rho^2}{32\sqrt{2\rho} \ n^{3/2}}+\dots
\,,
\ee
where there is in principle two possible behaviours depending on the signs $\pm$. Notice that the angular momentum only comes in the NNNLO correction, entering the $n^{-3/2}$ term.

\begin{figure}[!ht]
    \centering
    \begin{subfigure}{0.32\textwidth}
        \centering
        \includegraphics[width=\linewidth]{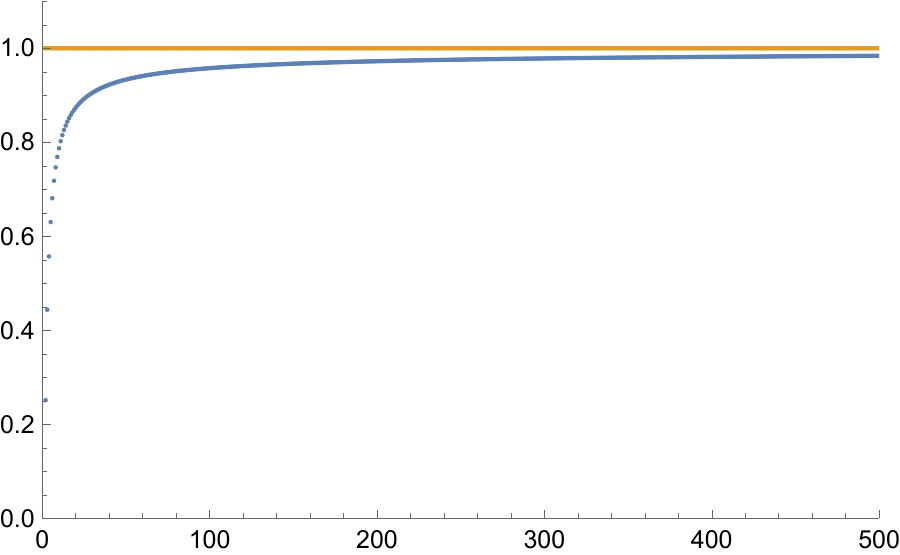}
        \caption{Fundamental mode}
    \end{subfigure}
    \hfill
    \begin{subfigure}{0.32\textwidth}
        \centering
        \includegraphics[width=\linewidth]{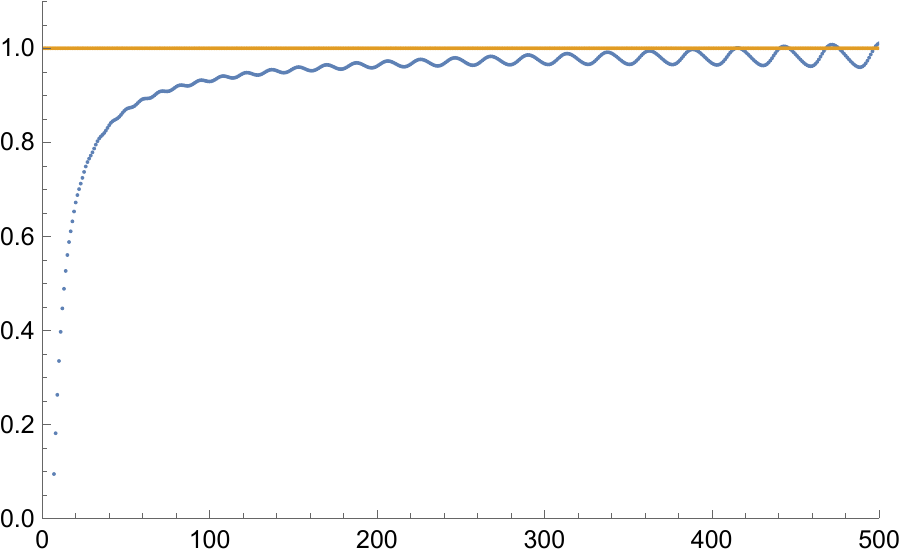}
        \caption{Fifth overtone}
    \end{subfigure}
    \hfill
    \begin{subfigure}{0.32\textwidth}
        \centering
        \includegraphics[width=\linewidth]{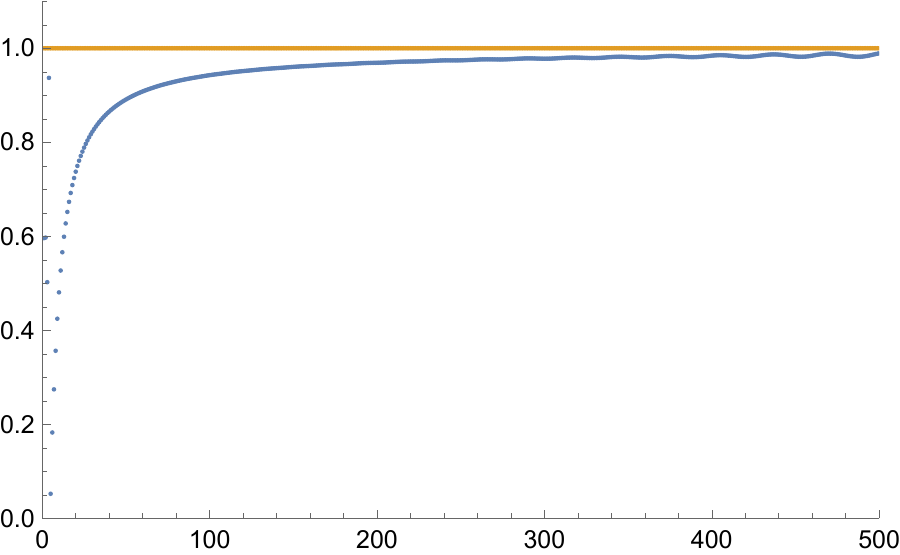}
        \caption{Random frequency}
    \end{subfigure}
    \caption{Examples of the behaviour of $R_n$ for different frequencies. We show plots of Re($R_n$) for the fundamental mode (on the left), the fifth mode (in the middle) and for a generic frequency ($0.2-2i$) here, for the Schwarzschild BH with $l=s=0$. Since Im($R_n$) vanishes in the limit $n\rightarrow\infty$, the modulus $|R_n|$ shows a behaviour similar to Re($R_n$). }
    \label{Rn_Sch_plot}
\end{figure}

Having a QNM solution necessarily imposes that the ratios approach 1 by below in modulus, thus selecting the $-$ branch (if one defines the square-root in the complex plane as always having a positive real part). It turns out that, as for now using our method, this criterion is not especially useful in identifying the QNM frequencies. Indeed, as illustrated on fig. \ref{Rn_Sch_plot}, depending on the value of the frequency, we have various behaviours.
One can see that the ratio $R_n$ converge to 1 for the fundamental mode, as expected, but also for random frequencies. Moreover a QNM frequency does not necessarily lead to a nice behaviour, since we see a growing oscillatory pattern for the fifth overtone for example. Thus the behaviour of $R_n$ cannot be used in a straightforward way to identify QNM frequencies in the complex plane.
%

Then the heart of Leaver's method is that the asymptotic QNM boundary condition at the horizon -- i.e. as $r$ approaches the Schwarzschild radius -- can be written in terms of continued fractions. Indeed, the ratios $R_n$ themselves can be expanded as continued fractions \cite{Leaver:1985ax}:
\be
\alpha_n R_{n+1}+\beta_n+\gamma_n \f1{R_{n}}=0
\quad\Rightarrow\quad
R_n
=\f{-\gamma_n}{\beta_n+\alpha_n R_{n+1}}
=\f{-\gamma_n}{\beta_n-\f{\alpha_n\gamma_{n+1}}{\beta_{n+1}+\alpha_{n+1} R_{n+2}} }
=\dots
\ee
Since the regularity at the horizon is that the field expansion is a well-defined power series, i.e. that the coefficient $a_{-1}$ vanishes, this condition translates
into a vanishing continued fraction equation:
\be
\alpha_0a_1+\beta_0a_0=0
\,\,\Rightarrow\,\,
0=
\beta_0+\alpha_0R_1
=
\beta_0-\f{\alpha_0\gamma_1}{\beta_1+\alpha_1R_2}
=
\beta_0-\f{\alpha_0\gamma_1}{\beta_1-\f{\alpha_1\gamma_2}{\beta_2+\alpha_2R_3}}
=
\beta_0-\f{\alpha_0\gamma_1}{\beta_1-\f{\alpha_1\gamma_2}{\beta_2-\f{\alpha_2\gamma_3}{\beta_3+\alpha_3R_4}}}
=\dots
\ee
Using the standard continued fraction notations, this can be written in a more compact fashion as:
\begin{equation}
\label{cont_frac}
    \beta_0-\frac{\alpha_0\gamma_1}{\beta_1-}\,\frac{\alpha_1\gamma_2}{\beta_2-}\,\frac{\alpha_2\gamma_3}{\beta_3-}...=0
    \,.
\end{equation}
This imposes the QNM boundary condition at the horizon.
A more compact form is obtained after factorizing the $\beta_{n}$'s allowing to rewrite the equation as a basic continued fraction with a unique set of coefficients $\zeta_n$:
\begin{equation}
\beta_0\left(1+\frac{\zeta_0}{1+}\,\frac{\zeta_1}{1+}\,\frac{\zeta_2}{1+}...\right)=0
\qquad\textrm{with}\quad
\zeta_{n}=-\f{\alpha_{n}\gamma_{n+1}}{\beta_{n}\beta_{n+1}}
\underset{n\to+\infty}{\to}-\f14\,.
\label{cont_frac_1coef}
\end{equation}
This standard form is simpler to handle and  benefits from the numerous theorems on continued fractions. 

Keeping in mind that all the coefficients $\alpha_n,\beta_n,\gamma_n$ are (rational) functions of the frequency $\om$, this approach sets an equation of which all the QNM frequencies must be zeros. In simpler words, the roots of this continued fraction equation give the QNM frequencies.
Nevertheless, as the continued fraction is an infinite expansion whose exact resummation is yet unknown, one  has to truncate it in order to determine these zeros numerically. Truncating it to a finite number of terms, say $\cN$, we would choose $\cN$ to optimize the numerical precision for the QNMs frequencies. In particular, as the imaginary part of the frequency grows, we must take larger values of $\cN$. 
To improve this, we use three extra ingredients.
First, one can invert the first terms of the continued fractions and send them to the other side of the equation. The $N$-th inversion of our equation \eqref{cont_frac} reads:
\be 
\beta_{N}-\frac{\alpha_{N-1}\gamma_{N}}{\beta_{N-1}-}\frac{\alpha_{N-2}\gamma_{N-1}}{\beta_{N-2}-}...-\frac{\alpha_{0}\gamma_{1}}{\beta_{0}}=\frac{\alpha_{N}\gamma_{N+1}}{\beta_{N+1}-}\frac{\alpha_{N+1}\gamma_{N+2}}{\beta_{N+2}-}\frac{\alpha_{N+2}\gamma_{N+3}}{\beta_{N+3}-}...
\,,
\ee
where we have a finite fraction on the left side of the equation and a continued fraction on the other.
As empirically observed in \cite{Leaver:1985ax}, this reformulation of the equation ensures a better stability and the $N$th inversion is apparently optimal  for the computation of the $N$th QNM. We therefore use this technique to improve the precision of our computations at high overtones.

Second, one can not brutally truncate the continued fraction. Setting the tail of the continued fraction to 0 creates numerical instabilities. One has to  approximate it instead. In fact, numerical convergence of continued fractions is the purpose of numerous mathematical studies, and computational methods are described thoroughly in~\cite{alma9917731286705941}.
Continued fractions with complex coefficients turn out to be rather subtle. The limit value $|\zeta_n|\rightarrow \f14$ turns out to be a critical value for their convergence. As long the sequence remains (asymptotically) below this threshold, one has an excellent stable control of the tail of the continued fraction\footnotemark{}.
\footnotetext{
Assuming the convergence of $\zeta_n \to \zeta$ as $n\to+\infty$,
we have an evaluation of the tail of the continued fraction:
\begin{equation}
\frac{\zeta}{1+}\,\frac{\zeta}{1+}\,\frac{\zeta}{1+}...=-\frac{1}{2}+\left(\frac{1}{4} +\zeta\right)^{1/2}\,,
\label{converg}
\end{equation}
which can be used to accelerate the convergence, as long as $|\zeta|\le \f14$. This argument can be extended to include $1/n$-corrections.
}
%
We will discuss the exact theorems and the derivation of analytical approximations to improve the numerical computation of complex continued fractions in a subsequent mathematical  paper.

Third, in order to determine the root values, we use extrapolations of the next root value as starting points for the root search. Simply guessing the next root at leading order is not particularly efficient and reaching frequencies with higher imaginary part requires a much finer ansatz, which we implement recursively\footnotemark{}.

\footnotetext{
Assuming for the n-th root an expansion of the form $\omega_n = \sum_{k=-m}^m c_{m/2} n^{m/2}$
where $m$ is a positive integer and $c_i$ complex coefficients, we obtain the extrapolations $\omega_{n+1}^{(m)}$ of order $m$ at large $n$:
\begin{eqnarray*}
    \omega_{n+1}^{(0)} &=& 2 \omega_{n} - \omega_{n-1} \\
    \omega_{n+1}^{(1)} &=& 3 \omega_{n} - 3 \omega_{n-1} + \omega_{n-2} \\
    \omega_{n+1}^{(2)} &=& 4 \omega_{n} - 6 \omega_{n-1} + 4 \omega_{n-2} - \omega_{n-3} \\
    \omega_{n+1}^{(3)} &=& 5 \omega_{n} - 10 \omega_{n-1} + 10 \omega_{n-2} - 5 \omega_{n-3} + \omega_{n-4} \\
    \omega_{n+1}^{(4)} &=& 6 \omega_{n} - 15 \omega_{n-1} + 20 \omega_{n-2} - 15 \omega_{n-3} + 6 \omega_{n-4} - \omega_{n-5}\\
     \omega_{n+1}^{(5)} &=& 7 \omega_{n} - 21 \omega_{n-1} + 35 \omega_{n-2} - 35 \omega_{n-3} + 21 \omega_{n-4} - 7 \omega_{n-5} + \omega_{n-6}
\end{eqnarray*}
The complete process to precisely compute the complex continued fractions and their roots will be described in a separate future paper focusing on mathematical aspects.
}

\medskip

Now that we have reviewed the basic tools in the standard Schwarzschild BH case, let us turn to BCL BH and see how to extract the Quasi-Normal frequencies from the five-term recursion relation \eqref{recur_eq}.
To start with, we can still translate the recursion relation for the $a_{n}$'s into a non-linear recursion for the ratio $R_{n}$:
\be
\alpha_n a_{n+1}+\beta_na_n+\gamma_na_{n-1}+\delta_na_{n-2}+\epsilon_na_{n-3}=0
\quad\Rightarrow\quad
\alpha_n R_{n+1}+\beta_n+\gamma_n \f1{R_{n}}
+\delta_n\f1{R_{n}R_{n-1}}+\epsilon_n\f1{R_{n}R_{n-1}R_{n-2}}
=
0
\,.
\ee
This implies the asymptotic behaviour for the large $n$ limit of the ratios:
\be
R_{n}
\underset{n\gg1}\sim
1\pm\sqrt{\f{2(1+r_-)\rho}{n}}+\f{2\rho-\f34}{n}+\frac{3+16\lambda-16(3+r_-)\rho+32(2-r_-)(1+r_-)\rho^2}{32\sqrt{2(1+r_-)\rho} \ n^{3/2}}+\dots
\,.
\ee
Actually, the recursion relation can allow for another behaviour, but our numerical exploration reveals that this case never occurs. 

Not only this higher order recursion relation does not allow to write the ratios $R_{n}$ as continued fractions, but the non-linearity of the recursion causes numerical instabilities.
In order to extract the Quasi-Normal frequencies from the five-term recursion relation \eqref{recur_eq}, one wishes to cast this recursion relation into a 3-term one, so that the continued fraction method can be used as in the Schwarzschild case \cite{Leaver:1985ax,Livine:2024bvo}.
This is possible by making use of the Gaussian reduction procedure, which allows to reduce step by step the depth of the recursion.
Starting with a first step defining new recursion coefficients,
\be
\bar{\alpha}_n= \alpha_n
\,,\quad
\bar{\beta}_n= \beta_n-\frac{\bar{\alpha}_{n-1}}{\bar{\delta}_{n-1}}\epsilon_n
\,,\quad
\bar{\gamma}_n= \gamma_n-\frac{\bar{\beta}_{n-1}}{\bar{\delta}_{n-1}}\epsilon_n
\,,\quad
\bar{\delta}_n= \delta_n-\frac{\bar{\gamma}_{n-1}}{\bar{\delta}_{n-1}}\epsilon_n
\,,
\ee
the five-term recursion relation is equivalent to a four-term recursion relation\footnotemark{},
\be
\bar\alpha_n a_{n+1}+\bar\beta_na_n+\bar\gamma_na_{n-1}+\bar\delta_na_{n-2}=0
\,.
\ee
\footnotetext{
The validity of the four-term recursion is proved by recursion. Indeed, plugging the definition of the new coefficients in terms of the old ones, we get for a given integer $n$:
\be
\bar\alpha_n a_{n+1}+\bar\beta_na_n+\bar\gamma_na_{n-1}+\bar\delta_na_{n-2}
=
(\alpha_n a_{n+1}+\beta_na_n+\gamma_na_{n-1}+\delta_na_{n-2})
-\f{\epsilon_n}{\bar{\delta}_{n-1}}(\bar\alpha_{n-1} a_{n}+\bar\beta_{n-1}a_{n-1}+\bar\gamma_{n-1}a_{n-2})
\,,
\nn
\ee
which vanishes, since the first term is equal to $-\eps_na_{n-3}$ due to the original 5-term recursion relation, and the second term is also equal to $-\eps_na_{n-3}$assuming that the new 4-term recursion relation holds for the integer $n-1$.
}
\ \\
Running this procedure a second time redefines new recursion coefficients:
\be
 \bar{\bar{\alpha}}_n= \bar{\alpha}_n
 \,,\quad
 \bar{\bar{\beta}}_n= \bar{\beta}_n-\frac{\bar{\bar{\alpha}}_{n-1}}{\bar{\bar{\gamma}}_{n-1}}\bar{\delta}_n
 \,,\quad
 \bar{\bar{\gamma}}_n= \bar{\gamma}_n-\frac{\bar{\bar{\beta}}_{n-1}}{\bar{\bar{\gamma}}_{n-1}}\bar{\delta}_n
 \,,
\ee
and we end up with a three-term recursion relation,
\be
\bar{\bar{\alpha}}_{n}a_{n+1}+\bar{\bar{\beta}}_{n}a_n+\bar{\bar{\gamma}}_{n}a_{n-1}=0.
\ee
The ratios $R_{n}$ then also satisfy a simpler relation,
\be
\bar{\bar{\alpha}}_n R_{n+1}+\bar{\bar{\beta}}_n+\bar{\bar{\gamma}}_n \f1{R_{n}}=0
\,,
\ee
and can now be expanded in continued fractions just as in the Schwarzschild BH case.
This is the method that we used here to compute the QNM frequencies for the BCL BH.
The difficulty with this algorithm is that the definition of  the new coefficients $\bar{\bar{\alpha}}_{n}, \bar{\bar{\beta}}_n, \bar{\bar{\gamma}}_n$  is itself recursive, and it is not possible to derive analytical expressions for them.
It is nevertheless possible to show that at leading order, we have the same \enquote{critical} behaviour as for Schwarzchild.
Indeed, we can compute the leading order behaviour of all the recursion coefficients. Starting with the original coefficients, we have:
\be
\left|\begin{array}{lcl}
\alpha_{n}&\sim& (r_{-}+1)n^{2}
\,,\\
\beta_{n}&\sim& 2(r_{-}+1)(r_{-}-1)n^{2}
\,,\\
\gamma_{n}&\sim& (r_{-}+1)(r_{-}-2-\sqrt{3})(r_{-}-2+\sqrt{3})n^{2}
\,,\\
\delta_{n}&\sim& -2r_{-}(r_{-}+1)(r_{-}-1)n^{2}
\,,\\
\eps_{n}&\sim& r_{-}^{2}(r_{-}+1) n^{2}
\,,
\end{array}
\right.
\ee
where we notice the special value $(2\pm\sqrt{3})$ for the deformation parameter $r_{-}$. Their physical relevance and effect on the BH phenomenology is not clear. Nonetheless, we can plug those leading order approximations in the Gaussian reduction and get two possible branches:
\be
\left|\begin{array}{lcl}
\bar{\alpha}_{n}&\sim& (r_{-}+1)n^{2}
\,,\\
\bar{\beta}_{n}&\sim& (r_{-}+1)(r_{-}-2)n^{2}
\,,\\
\bar{\gamma}_{n}&\sim& -(r_{-}+1)(2r_{-}-1)n^{2}
\,,\\
\bar{\delta}_{n}&\sim& r_{-}(r_{-}+1)n^{2}
\end{array}
\right.
\quad\textrm{or}\qquad
\left|\begin{array}{lcl}
\bar{\alpha}_{n}&\sim& (r_{-}+1)n^{2}
\,,\\
\bar{\beta}_{n}&\sim& (r_{-}+1)(2r_{-}-1)n^{2}
\,,\\
\bar{\gamma}_{n}&\sim& (r_{-}+1)r_{-}(r_{-}-2)n^{2}
\,,\\
\bar{\delta}_{n}&\sim& -r_{-}^{2}(r_{-}+1)n^{2}
\,.
\end{array}
\right.
\ee
Repeating the process for the second Gaussian reduction leads to an extra possible branches for the asymptotic behaviour:
\be
\left|\begin{array}{lcl}
\bar{\bar{\alpha}}_{n}&\sim& (r_{-}+1)n^{2}
\,,\\
\bar{\bar{\beta}}_{n}&\sim& -2(r_{-}+1)n^{2}
\,,\\
\bar{\bar{\gamma}}_{n}&\sim& (r_{-}+1)n^{2}
\,,
\end{array}
\right.
\quad\textrm{or}\qquad
\left|\begin{array}{lcl}
\bar{\bar{\alpha}}_{n}&\sim& (r_{-}+1)n^{2}
\,,\\
\bar{\bar{\beta}}_{n}&\sim& (r_{-}+1)(r_{-}-1)n^{2}
\,,\\
\bar{\bar{\gamma}}_{n}&\sim& -r_{-}(r_{-}+1)n^{2}
\,,
\end{array}
\right.
\quad\textrm{or}\qquad
\left|\begin{array}{lcl}
\bar{\bar{\alpha}}_{n}&\sim& (r_{-}+1)n^{2}
\,,\\
\bar{\bar{\beta}}_{n}&\sim& 2r_{-}(r_{-}+1)n^{2}
\,,\\
\bar{\bar{\gamma}}_{n}&\sim& r_{-}^{2}(r_{-}+1)n^{2}
\,.
\end{array}
\right.
\ee
They are only two possible limit for the corresponding $\zeta_{n}$ coefficients:
\be
\zeta_{n}=-\f{\bar{\bar{\alpha}}_{n}\bar{\bar{\gamma}}_{n+1}}{\bar{\bar{\beta}}_{n}\bar{\bar{\beta}}_{n+1}}
\underset{n\to+\infty}{\to}-\f14
\quad\textrm{or}\quad
\f{r_{-}}{(r_{-}+1)^{2}}
\,.
\ee
Surprisingly, our numerical computations reveal that the latter value, which depends on $r_{-}$, seems never to be realized, thereby implying that the asymptotic behaviour of the continued fraction is similar to the Schwarzschild case.
Nevertheless, despite this leading order analysis, we have not found any closed exact formula for the coefficients $\bar{\bar{\alpha}}_{n}, \bar{\bar{\beta}}_n, \bar{\bar{\gamma}}_n$. It is thus necessary for us to evaluate them numerically, which leads to an extra layer of numerical errors which we need to keep under control, on top of the subtleties already mentioned for the Schwarzschild case.

\section{Analytical computations for the asymptotic part of the spectrum}
\label{Analytic_section}

This section is dedicated to analytical predictions for the spectrum of high overtone QNMs, which will then be compared with the numerical computations. 
We use the monodromy technique based the properties of the field around the metric singularities at the horizon and at $r=0$ in the complex plane, e.g. \cite{Motl:2003cd}, and the near-horizon $\sl_2$ symmetry e.g. \cite{Bertini:2011ga}. Another possible method would have been to study the convergence rate of the continued fractions, as in \cite{Motl:2002hd}, but we haven't been able to adapt it to the more complicated high order recursion relation of the BCL black hole case, due to the lack of analytical formulas for the recursion coefficients $\bar{\bar{\alpha}}_{n}, \bar{\bar{\beta}}_n, \bar{\bar{\gamma}}_n$ after (double) Gaussian reduction.

Even if the high overtones are not a priori the most observationally useful QNMs, they are especially relevant for two reasons.
First, in practice, this asymptotic tail is simpler mathematically: it is the part of the QNM spectrum over which we have the best analytical control (besides the first few modes which can be computed using the WKB approximation).
Second, while those highly dissipative modes are not the dominant drive for the ringdown of binaries, they have been argued to reflect the deep quantum regime of black holes and reveal the quantum gravity Planck-scale fluctuations, e.g. \cite{Dreyer:2002vy}. It is in this yet uncharted regime that we expect the largest deviation from general relativity. As we shall see later, it is indeed in this regime that we will witness a large systematic deviation between the test-field QNMs and the physical perturbation QNMs for the BCL black holes, which is the main result of the present paper.

\subsection{Analytical computation of the asymptotics using the monodromy technique}

In this section we will focus more specifically on the large imaginary part of the spectrum where the QNMs are highly damped.
It is known that, for the Schwarzschild BH, the real part of the QNM frequencies converges towards the constant value $(\ln 3)/4\pi$ while the gap in imaginary part between two succeeding QNMs converges to a value of around $\frac{1}{2}$: 
\be 
\boxed{4\pi r_s \omega =\ln(3)-i(2n+1)\pi, \ \ \ \ n\in\mathbb{N}.}
\label{Sch_mono}
\ee 
The gap between the QNMs imaginary parts is supposed to probe the deep quantum regime of BHs. See (\cite{Motl:2002hd,Motl:2003cd,Natario:2004jd,Dreyer:2002vy,Konoplya:2011qq}) for reviews. 
Here, we want to use the same technique of computation which was used to determine \eqref{Sch_mono}, referred as the monodromy technique, to probe the asymptotics of the test-field QNM spectrum of the BCL BH and see how the parameter $r_-$ impacts equation \eqref{Sch_mono}.
\\ \\
Let us consider the effective potential for spin 0 and 2 test-field perturbations of the BCL BH
\be 
V_s=\frac{f(r)}{r}\lp\frac{\lambda}{r}+(1-s^2)\frac{r_+-r_-}{r^2}+(2-3s)\frac{r_-r_+}{r^3}\rp,
\label{BCLs02pot}
\ee
along with the equation 
\be 
\frac{\dd^2 \Phi_s}{\dd x^2} + [\omega^2-V_s(r)]\Phi_s = 0.
\label{schro2}
\ee
The key point of the monodromy technique is to extend the usual physical domain of definition $r_+<r<\infty$ to the hole complex plane. Then, the equation \eqref{schro2} is an ordinary differential equation with regular singular points at $r=0$, $r=r_+$, $r=-r_-$ and an irregular singular point at $r=\infty$. Thus, any solution from \eqref{schro} in the physical region extends to the complex plane, but may be multivalued around the singular points. 
\\ \\

Let us first take a look at infinity. As the effective potential \eqref{BCLs02pot} vanishes at infinity, the wave-function $\Phi$ has a plane-wave behaviour:
\be 
\Phi(x)\overset{\infty}{\sim} A_+e^{i\omega x}+A_-e^{-i\omega x}
\ee 
The boundary condition for QNMs at infinity, stating that the wave must be outgoing at infinity, implies $A_-=0$ and then $\Phi(x)\overset{\infty}{\sim} A_+e^{i\omega x}$.
\\ \\

Now let us consider the behaviour of the wave-function $\Psi$ around the singularity $r=0$. At first order, the effective potential \eqref{BCLs02pot} reduces to 
\be 
V_s \overset{r=0}{\sim} (3s-2)\frac{r_-^2r_+^2}{r^6}.
\label{1stPotr}
\ee 
It is important to notice that this potential vanishes when taking $r_-$ to 0 and that we do not recover the first order Schwarzschild potential $V_{\text{Sch}} \overset{r=0}{\sim}\frac{3r_s^2}{r^4}$, which is the one considered in the monodromy technique for the Schwarzschild BH. As we will see later, this will result in the lack of a well defined limit $r_-\longrightarrow 0$ for the monodromy result and its prediction is then expected to be less accurate for small $r_-$. One would need to go beyond the first order to overcome this issue.
\\
The first order of the tortoise coordinate around the singularity $r=0$ writes
\be 
x\overset{r=0}{\sim} -\frac{r^3}{3r_-r_+}.
\label{x_at_r0}
\ee 
Notice once again that the limit $r_-\longrightarrow0$ is not well define.\\
The potential \eqref{BCLs02pot} can then be written as a function of $x$:
\be 
V_s(x)=\frac{4}{9x^2}.
\label{1stPotx}
\ee 
The equation \eqref{schro2} for spin 0 and 2 test-field perturbations around the BCL BH can thus be simplified around $r=0$ as
\be 
\frac{\dd^2 \Phi}{\dd x}^2 + \lp\omega^2-\frac{4}{9x^2}\rp \Phi  =0.
\ee
It is convenient to rescale the $x$-coordinate into $z=\om x$ and introduce the function $\Psi$ such that $\Phi(x)=\Psi(\omega x)\,{\sqrt{2\pi\omega x}}$. This function now satisfies a straightforward Bessel equation:
%
\be 
z^2\pp_z^2\Psi+z\pp_z\Psi+\lp z^2-\nu^2\rp\Psi=0,
\qquad\textrm{with}\quad
\nu=\frac{\sqrt{1+12s}}{6}.
\ee 
Thus the general solution of equation \eqref{1stPotx} for the wave function $\Phi(x)$ around the singularity $r=0$ can be written in terms of Bessel functions of the first kind $J_{\pm\nu}$:
\be 
\Phi(x) \overset{r=0}{\sim} B_+\sqrt{2\pi \omega x} \ J_\nu(\omega x) + B_-\sqrt{2\pi\omega x} \ J_{-\nu}(\omega x)
\label{solB}
\ee

\begin{figure}[!ht]
    \centering
    \includegraphics[width=0.5\linewidth]{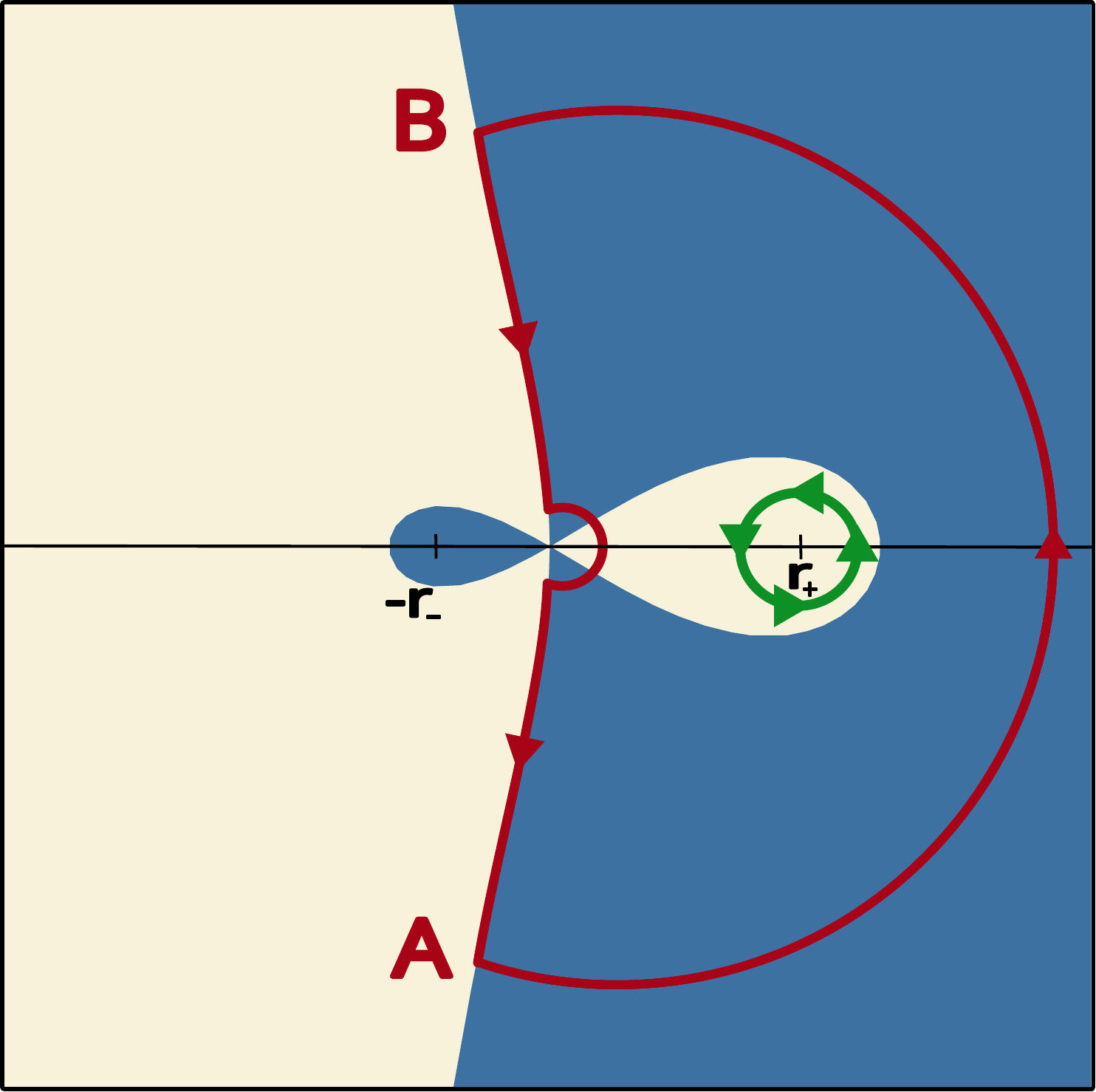}
    \caption{Contour for the calculation of QNM frequencies in the complex $r$ plane. The different colour regions are separated by the associated Stokes lines and the dark blue region corresponds to Re$(x)>0$.}
    \label{BCL_r_plane}
\end{figure}

The next thing we want to do is matching this solution 
with the solution at infinity $\Phi(x)\overset{\infty}{\sim} A_+e^{i\omega x}$. In this section we are interested in the study of the asymptotic part of the QNM spectrum. Whereas considering the Schwarzschild BH or the BCL BH from modified gravity, one see that going through the spectrum the absolute value of the imaginary part of the QNMs $|\mathrm{Im}(\omega)|$ soon becomes way larger than the real part $\mathrm{Re}(\omega)$. In the asymptotic part of the spectrum, we can then consider $\omega$ to be very large and purely imaginary. We will stick to this hypothesis throughout all the computation.\\
We can now use the following expansion of the Bessel functions
\be
J_\nu(z)\overset{z\gg1}{\sim} \sqrt{\frac{2}{\pi z}} \cos\lp z-\frac{\nu \pi}{2}-\frac{\pi}{4}\rp,
\ee 
and apply it to the solution \eqref{solB} around $r=0$ for $|\omega x|\gg 1$:
\begin{subequations}
    \begin{align}
        &\Phi(x)\overset{|\omega x|\gg1}{\sim} 2B_+\cos\lp \omega x-\frac{\nu \pi}{2}-\frac{\pi}{4}\rp +  2B_-\cos\lp \omega x+\frac{\nu \pi}{2}-\frac{\pi}{4}\rp,\\
        \iff &\Phi(x)\overset{|\omega x |\gg1}{\sim} \lp B_+ e^{-i\alpha_+}+B_- e^{-i\alpha_-}\rp e^{i\omega x}+ \lp B_+ e^{i\alpha_+} + B_- e^{i\alpha_-}\rp e^{-i\omega x},
\label{BCL_solu1}
    \end{align}
\end{subequations}
with $\alpha_\pm=\frac{\pi}{4}(1\pm 2\nu)$.
In order to  match  the solutions at infinity and  around $r=0$, none of the exponentials $e^{\pm i\omega x}$ should dominate each other. 
The matching with the solution at infinity must then be done following the Stokes lines, defined such that $x$ is purely imaginary and then such that $\omega x \in \mathbb{R}$. The Stokes lines are represented in black in the complex r-plane of figure \ref{BCL_r_plane}. For the calculation of the QNMs frequencies we will follow the red contour, starting at point $B$ where $\mathrm{Re}(\omega x)>0$. Indeed, we have $\mathrm{Im}(\omega)<0$ as we use the convention $e^{-i\omega t}$ for the time evolution, along with $\mathrm{Im}(x)$ being positive at point $B$.
 
As $\omega x \gg 1$ at point $B$, we can use the solution \eqref{BCL_solu1} and apply the QNM boundary condition at infinity, \textit{i.e.} the condition that the wave must be outgoing at infinity. This give us a first condition: 
\be 
B_+ e^{i\alpha_+} + B_- e^{i\alpha_-} = 0,
\label{first_cond}
\ee 
and we are left with the following solution for the wave function $\Phi(x)$
\be 
\Phi(x)\overset{\omega x \gg 1}{\sim} \lp B_+ e^{-i\alpha_+}+B_- e^{-i\alpha_-}\rp e^{i\omega x}.
\ee 
In order to rotate from the branch containing the point B to the one containing the point A, we need to rotate around the singularity $r=0$ by an angle $-\pi$ in the complex $r$-plane. From equation \eqref{x_at_r0}, one can see that this is equivalent to rotate by angle $-3\pi$ in the $x$-plane.
Let us apply this to our solution, using the asymptotic expansion for the Bessel function under a $-3\pi$ rotation:
\begin{subequations}
    \begin{align}
        \sqrt{2\pi e^{-3\pi i}\omega x} \ J_{\pm \nu}(e^{-3\pi x}\omega x)&= e^{-\frac{3}{2}\pi i\lp 1\pm 2\nu \rp}\sqrt{2\pi \omega x} \ J_{\pm\nu}(\omega x)\\
        &\sim 2 e^{-6i\alpha_\pm}\cos(\omega x -\alpha_\pm).
    \end{align}
\end{subequations}
The solution at the point A can then be written as:
\begin{subequations}
    \begin{align}
        \Phi(x)&\sim 2 B_+ e^{-6i\alpha_+}\cos(-\omega x -\alpha_+) + 2 B_- e^{-6i\alpha_-}\cos(-\omega x -\alpha_-)\\
        &= (B_+e^{-7i\alpha_+}+B_- e^{-7i\alpha_-})e^{-i\omega x} + (B_+e^{-5i\alpha_+}+B_- e^{-5i\alpha_-})e^{i\omega x}.
    \end{align}
\end{subequations}
Finally, let us close the contour around $r\sim\infty$, following the red line, where $x\sim r$ and Re$(x)>0$. As Im$(\omega)\ll 0$, the factor $e^{-i\omega x}$ is exponentially  small on this part of the contour, so that only the coefficient of $e^{i\omega x}$ should be taken into account for the calculation. After having completed the red contour, the coefficient $e^{i\omega x}$ is then multiplied by 
\be 
\frac{B_+e^{-5i\alpha_+}+B_- e^{-5i\alpha_-}}{B_+e^{-i\alpha_+}+B_- e^{-i\alpha_-}}.
\ee 
Moreover, as $x\overset{r_+}{\sim}\frac{1}{f'(r_+)}\log(r-r_+),$ the counterclockwise monodromy of $e^{i\omega x}$ is
\be 
e^{i\omega \lp\frac{1}{f'(r_+)}2\pi i\rp}= e^{i\omega \times 2\pi i \frac{r_+^2}{r_-+r_+}}= e^{-2\pi \omega \frac{r_+^2}{r_-+r_+}},
\ee 
so that the counter-clockwise monodromy of the wave function $\Phi$ around this contour is 
\be 
\frac{B_+e^{-5i\alpha_+}+B_- e^{-5i\alpha_-}}{B_+e^{-i\alpha_+}+B_- e^{-i\alpha_-}}e^{-2\pi \omega \frac{r_+^2}{r_-+r_+}}.
\ee 
One can distort the contour without any impact on the monodromy value, as long as the distortion does not cross any singularity. Let us then take a contour simply circling the only singularity inside, being $r_+$. This contour is drawn in figure \ref{BCL_r_plane} as the small green circle.
As the effective potential \eqref{BCLs02pot} vanishes around the horizon $r_+$, the wave-function $\Phi$ has a plane-wave shape:
\be 
\Phi(x)\overset{r_+}{\sim} C_+e^{i\omega x}+C_-e^{-i\omega x}.
\ee 
We can apply the QNM boundary condition at the horizon, stating that the wave must be ingoing. This implies $C_+=0$ and then
\be 
\Phi(x)\overset{r_+}{\sim} C_-e^{-i\omega x}.
\ee 
The counter-clockwise monodromy of the wave function $\Phi$ around the green contour is then
\be 
e^{-i\omega \lp\frac{1}{f'(r_+)}2\pi i\rp}= e^{2\pi \omega \frac{r_+^2}{r_-+r_+}}.
\ee 
As the monodromy for the red contour should be equal to the monodromy for the green contour, we obtain a second condition in addition to the first one \eqref{first_cond}:
\be 
\frac{B_+e^{-5i\alpha_+}+B_- e^{-5i\alpha_-}}{B_+e^{-i\alpha_+}+B_- e^{-i\alpha_-}}e^{-2\pi \omega \frac{r_+^2}{r_-+r_+}} = e^{2\pi \omega \frac{r_+^2}{r_-+r_+}}.
\label{secd_cond}
\ee 
At the end of the day, we have two conditions, equations \eqref{first_cond} and \eqref{secd_cond}, representing the two QNM boundary conditions, at infinity and at the horizon. They form a system of 2 equations with 2 variables $B_\pm$. In order to have solutions, the determinant of the system must vanish: 
\be
\left|
\begin{array}{cc}
e^{i\alpha_+} & e^{i\alpha_-}\\
e^{-5i\alpha_+}e^{-2\pi \omega R}-e^{-i\alpha_+}e^{2\pi \omega R} \ \ \  &  \ \ \ e^{-5i\alpha_-}e^{-2\pi \omega R}-e^{-i\alpha_-}e^{2\pi \omega R}
\end{array}
\right|=0,
\ee 
where we have introduced a characteristic length scale:
\be 
R= \frac{r_+^2}{r_-+r_+}.
\ee 
The determinant equation expands as:
\be
e^{-2\pi \omega R}\lp e^{i(\alpha_+-5\alpha_-)}-e^{i(\alpha_--5\alpha_+)}\rp  + e^{2\pi \omega R}\lp e^{i(\alpha_--\alpha_+)}-e^{i(\alpha_+-\alpha_-)}\rp =0.
\ee
Plugging the explicit values for the angles $\alpha_\pm$, this equation reduces to
\be 
e^{4\pi \omega R} = -\frac{\sin(3\pi \nu)}{\sin(\pi\nu)} = -2,
\ee 
\
as $\nu=\frac{1}{6}$ for $s=0$ and $\nu=\frac{5}{6}$ for $s=2$.

We finally obtain a condition for the QNMs frequencies $\omega$ of spin $s=0,2$ test field perturbations on a BCL BH in the high-imaginary part of the spectrum: 
\be 
\boxed{4\pi R \omega =\log(2)-i(2n+1)\pi, \ \ \ \ n\in\mathbb{N}.}
\label{mono_BCL}
\ee 
This condition for asymptotic test-field QNMs of BCL has a very similar shape to the asymptotic QNMs of a Schwarzschild BH $\eqref{Sch_mono}$. Yet we can identify two major differences. The first one is the length scale factor  $R = \frac{r_+^2}{r_-+r_+}$, implying that the frequencies explicitly depend on the inner radius $r_-$ in addition to the exterior horizon $r_+$. In the limit $r_- \rightarrow 0$, the factor $R$ reduces to the Schwarzschild radius $r_s$. The second major disparity is the presence of a $\log(2)$ in the real part of the \eqref{mono_BCL} equation, instead of a $\log(3)$ as in the Schwarzschild equation \eqref{Sch_mono}. This implies that the limit $r_- \rightarrow 0$ of equation \eqref{mono_BCL} does not give the Schwarzschild monodromy condition \eqref{Sch_mono} and that there is no continuity between those equations. This issue directly comes from the restriction on the first order expressions for the effective potential \eqref{1stPotr} and the tortoise coordinate \eqref{x_at_r0} for the monodromy computation. Indeed, we had already highlighted the lack of well-defined limit $r_-\rightarrow 0$ for these expression. In order to obtain a monodromy result for the QNM asymptotic frequencies with a good Schwarzschild limit, one would need to go beyond the first order in equations \eqref{1stPotr} and \eqref{x_at_r0} -- up to the order for which a Schwarzschild limit is well-defined -- which would result in a much more involved computation.\\
We then expect the monodromy result \eqref{mono_BCL} to be less accurate for small values of $r_-$. We will examine this question in section \ref{Comp_Asymp} where we will compare our monodromy analytical result with the QNMs numerically computed thanks to the Leaver method.

\subsection{Asymptotic spectrum from near-horizon sl${}_{2}$ symmetry}

%

Another approach for extracting the asymptotic QNM frequencies is to exploit the $\sl_{2}$ symmetry of the near-horizon physics. The near-horizon approximation, developed for instance in \cite{Bertini:2011ga}, is based on the approximation of the potential term at leading order. It reveals a near-horizon $\sl_{2}$ symmetry for test fields evolving in Schwarzschild backgrounds and other black hole metrics. We show here that this logic applies to BCL black holes and allows to derive in a very simple fashion the gap between frequencies for the large overtone asymptotic.

This derivation of asymptotic frequencies somewhat clashes with the line of thoughts underlying the monodromy calculations. Indeed, on the one hand, the approach by monodromy reflects the whole BH space-time, crucially relies on the QNM boundary condition at both horizon and spatial infinity, and further depends on the metric at the singularity within the BH (at $r=0$ for Schwarzschild-like metrics). On the other hand, the near-horizon symmetry only relies on the near-horizon metric, as its name properly conveys. 
%
%
The contrast between the two logics and their assumptions hints towards a deeper reason for the structure of the frequencies in the highly dissipative regime and a more fundamental physical and mathematical mechanism protecting the asymptotic spectrum.

\medskip

Let us start with the scalar field, with spin $s=0$.
We recall the BCL space-time metric
\be
\dd s^2=-f(r)\dd t^2+\frac{1}{f(r)}\dd r^2+r^{2}\dd \Omega^2,
\qquad\textrm{with}\quad
f(r)=r^{-2}\sigma(r)
\quad\textrm{and}\quad
\sigma(r)=(r-r_{+})(r+r_{-})
\,.
\ee
The Klein-Gordon equation for a massless scalar field $\Phi_0=e^{-i\om t}\phi(r)Y^{l}_{m}(\theta,\vphi)$ reads:
\be
\f1{\sqrt{-g}}\pp_{\mu}\left(
\sqrt{-g}\,g^{\mu\nu}\pp_{\nu}\Phi_0
\right)
=0
\quad\Leftrightarrow\quad
\pp_{r}(\sigma\pp_{r}\phi)+\f{\om^{2}r^{4}}\sigma\phi=l(l+1)\phi
\,.
\ee
The near-horizon approximation consists in rounding off the frequency pre-factor at leading order around the horizon $r\sim r_{+}$:
\be
\f{\om^{2}r^{4}}\sigma
\underset{r\sim r_{+}}\sim
\f{\om^{2}r_{+}^{4}}\sigma
\,.
\ee
This leads to a modified approximate Klein-Gordon equation:
\be
\pp_{r}(\sigma\pp_{r}\phi)-\f{r_{+}^{4}}\sigma\pp_{t}^{2}\phi=l(l+1)\phi
\,.
\ee

It turns out that the modified Klein-Gordon differential operator can be written as the quadratic Casimir of the $\sl_{2}$ Lie algebra formed by the following three vector fields:
\be
L_{0}=+2R\pp_{t}
\,,\quad
L_{\pm}=
\pm \f{1}{\sqrt{\sigma}}e^{\pm \f t{2R}}\,\Big{[}
\sigma \pp_{r}\mp R\pp_{r}\sigma\, \pp_{t}
\Big{]}
\,,
\ee
with the same convention than in the previous sections, $R=r_{+}^{2}/(r_{+}+r_{-})$.
These first order differential operators form a $\sl_{2}$ algebra:
\be
[L_{0},L_{\pm}]=\pm L_{\pm}\,,\quad
[L_{+},L_{-}]=-2 L_{0}\,.
\ee
Then the approximate Klein-Gordon operator is simply the $\sl_{2}$ Casimir:
\be
\pp_{r}(\sigma\pp_{r}\phi)-\f{r_{+}^{4}}\sigma\pp_{t}^{2}\phi=l(l+1)\phi
\,\,\Leftrightarrow\,\,
\hcC\,\phi\,=\,l(l+1)\phi
\qquad\textrm{with}\quad
\hcC
\equiv
L_{0}^{2}
-\f12\left(
L_{-}L_{+}+L_{+}L_{-}
\right)
\,.
\ee
Since the $\sl_{2}$ operators  commute with their Casimir by definition, they act on the space of solutions, meaning that, if $\phi$ is a solution to the modified Klein-Gordon equation, then $L_{a}\phi$, with $a=0,\pm$, is also a solution to this equation.
We can thus use the algebraic toolkit of the representation theory of $\sl_{2}$. Following \cite{Bertini:2011ga}, we construct the QNMs as a highest weight representation built from a ``vacuum'' state diagonalizing $L_{0}$ and annihilated by $L_{+}$.
However, an important point to keep in mind is that the quasi-normal modes do not correspond to unitary representations. In fact, we do not define any Hermitian product and we do not identify the QNMs as orthonormal basis states (see e.g. for an approach to this question \cite{Green:2022htq}).

The eigenvalue of $L_{0}=2r_{s}\pp_{t}$ gives the frequency of the field configuration:
\be
L_{0}\phi^{(0)}
=
-2iR\om\, \phi^{(0)}
\,,
\ee
while the highest weight condition gives its precise dependency in the radial coordinate:
\be
L_{+}\phi^{(0)}=0
\,\,\Leftrightarrow\quad
\pp_{r}\phi^{(0)}
=
\f{\pp_{r}\sigma}{\sigma}\pp_{t}\phi^{(0)}
=
-i\om\,R\,\f{\pp_{r}\sigma}{\sigma}\phi^{(0)}
\,\,\Leftrightarrow\quad
\phi^{(0)}\propto e^{-i\om R\ln\sigma}
\,.
\ee
This behaves exactly as expected near the horizon:
\be
\ln\sigma
\underset{r\sim r_{+}}\sim
\ln(r-r_{+}) + \text{cst}
\quad \Rightarrow\quad
\phi^{(0)}
\underset{r\sim r_{+}}\propto
e^{-i\om}R\ln(r-r_{+}) 
\,,
\ee
appropriately fitting the required QNM boundary condition \eqref{boundaryhorizon} at the horizon.
Moreover, imposing that the mode is not (exponentially) suppressed at the horizon (as $x \rightarrow -\infty$) amounts to requiring a negative imaginary part of the frequency, $\textrm{Im}(\om)<0$.
We do not check the asymptotic boundary condition at spatial infinity $r\rightarrow+\infty$ since we are working in a near-horizon approximation, which can not ensure the proper behaviour at radial infinity.

Computing the Casimir $\hcC$ on  this highest weight mode fixes the value of its frequency in terms of the angular momentum $l$:
\be
l(l+1)\phi^{(0)}
=
\hcC\phi^{(0)}
=
L_{0}\left(
L_{0}+1
\right)\phi^{(0)}
=
-2iR\om(-2iR\om+1)\,\phi^{(0)}\,,
\ee
with two solutions:
\be
\om_{0}=\f i{2R}l
\quad\textrm{or}\quad
\om_{0}=-\f i{2R}(l+1)
\,.
\ee
We select the latter branch $2R\om=-i(l+1)$, with negative imaginary value, in order to satisfy the QNMs' boundary conditions.
Then higher modes are constructed by applying the $\sl_{2}$ generator $L_{-}$ as creation operator:
\be
\phi^{(n)}\propto (L_{-})^{n}\phi^{(0)}
\quad\Rightarrow\quad
L_{0}\phi^{(n)}
=-2iR\om_{n}\, \phi^{(n)}
=\left(-2iR\om_{0}-n\right)\, \phi^{(n)}\,,
\ee
thus giving an equally-spaced asymptotic spectrum along the negative imaginary semi-axis:
\be
2R\om_{n}=2R\om_{0}-in
\,.
\ee
This fits perfectly with the asymptotic spectrum prediction of the monodromy calculation, up to the precise value of the base frequency $\om_{0}$. Obtaining the correct real part would most certainly require a finer analysis, pushing for instance the near-horizon approximation to the next-to-leading order.
Actually, the near-horizon approximation, as reviewed here, is such a crude 0-th order truncation of the potential, that we can not truly trust the value $\om_0$ but can only consider as robust the predicted scaling of the frequency $\om$ in the mode label $n$.
In fact, at this approximation level, we obtain exactly the same prediction for the asymptotic QNM frequency at high overtone for test fields with spin $s=2$.

While it might be disappointing that this approximation does not achieve a better control over the QNM spectrum, it nevertheless underlines the robustness of the purely imaginary gap $\Delta \om=-i/2R$ between successive QNM frequencies for high overtones.
This is indeed here that we will see the most blatant deviation for QNMs between test-fields and physical perturbations.

\section{Numerical results from the Continued Fraction Method}

\subsection{The numerical process}

In this section we will present our main numerical results for the spin $s=0$ and $s=2$ test-field QNMs of the BCL BH. The Quasi-Normal frequencies were obtained by the application of the Continued Fraction method -- detailed in subsection \ref{Leaver_section} -- to the case of BCL's $s=0,2$ spin field perturbations. As we said before, this numerical method is one of the most accurate one existing and allows for the exploration of a large part of the QNM spectrum. This method does, however, have its own limitations and has turned out to be much more difficult to apply than in the Schwarzschild case. \\

The low damped mode are easily found as roots of the continued fraction \eqref{cont_frac} or of its first inversions, but the more we go down along the imaginary axis, the more difficult it is to numerically find those roots. 
In this work, we have been able to take the QNM search down to -20i or -50i -- depending on the value of $r_-$ --, which still allowed us to analyse the asymptotic part, as we will see in section \ref{Comp_Asymp}.\\
We explain these numerical difficulties by the two additional terms in the recursion equation \eqref{recur_eq} in comparison with the 3-terms Schwarzschild one. It is then required to numerically performed the Gauss reduction for each frequency, which inevitably results in increased computation times and a higher risk of numerical errors. That being said, we have ensured the validity of our results by comparing our low-damped results with a WKB approximation (see appendix \ref{WKB_appendix}), by tracking the evolution of the QNM's real parts and imaginary part when going lower along the imaginary axis, and by comparing the high-damped results with the monodromy prediction \eqref{mono_BCL}.
\\

As we stated in our previous paper \cite{Livine:2024bvo}, the Continued Fraction method is unreliable for purely imaginary frequencies (\textit{i.e.} purely damped modes), as many spurious modes are found along the imaginary axis. As we will see, the $s=2$ test-field QNMs cross the imaginary axis in the same way as the Schwarzschild QNMs. Yet, in this paper, we do not take a position on the presence or absence of purely imaginary QNMs for test-field BCL QNMs. The reader can still notice that we were able to identify a QNM around -10.4$i$ for $s=2$ and $l=3$ (see figure \ref{QNM_l}) with a very low real part: $\pm 0.00318$.
\\

In the following subsection we will show an overview of the main characteristics of the QNM spectra for test-fields perturbations of a BCL BH. Then in the last subsection we will focus on the asymptotic part of the QNM spectra in order to compare our analytical results from section \ref{Analytic_section} with our numerical section. The comparison between our $s=2$ test-field QNMs results with the physical QNMs computed in \cite{Roussille:2023sdr} can be found in the next section.
\\
We will set $r_+=1$ for all the numerical results. 

\subsection{Overview of the results}

%
\begin{figure}[htbp]
\centering

\begin{subfigure}{0.73\textwidth}
        \centering
        \includegraphics[width=.85\linewidth]{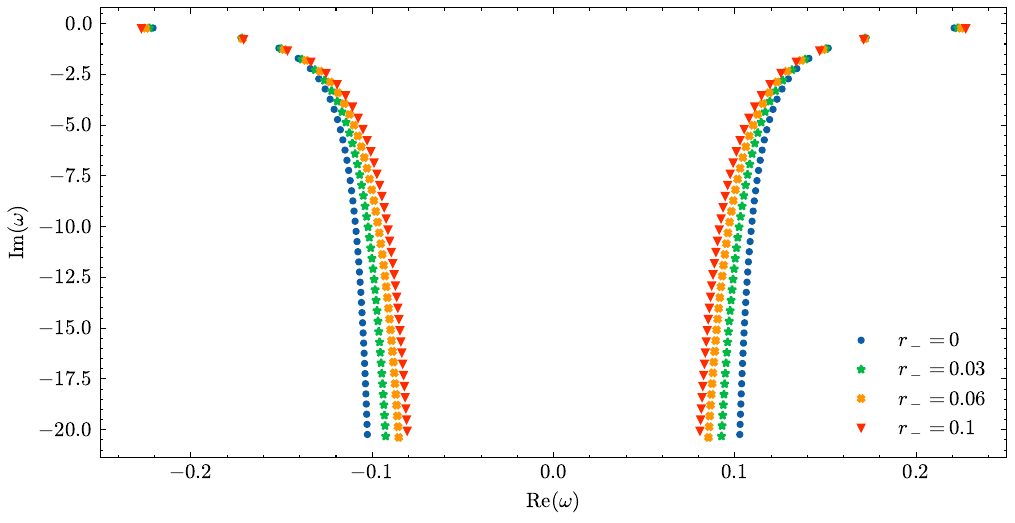}
        \caption{$r_-$ ranging from 0 to 0.1.}
    \end{subfigure}\\
    \begin{subfigure}{0.73\textwidth}
        \centering
        \includegraphics[width=.85\linewidth]{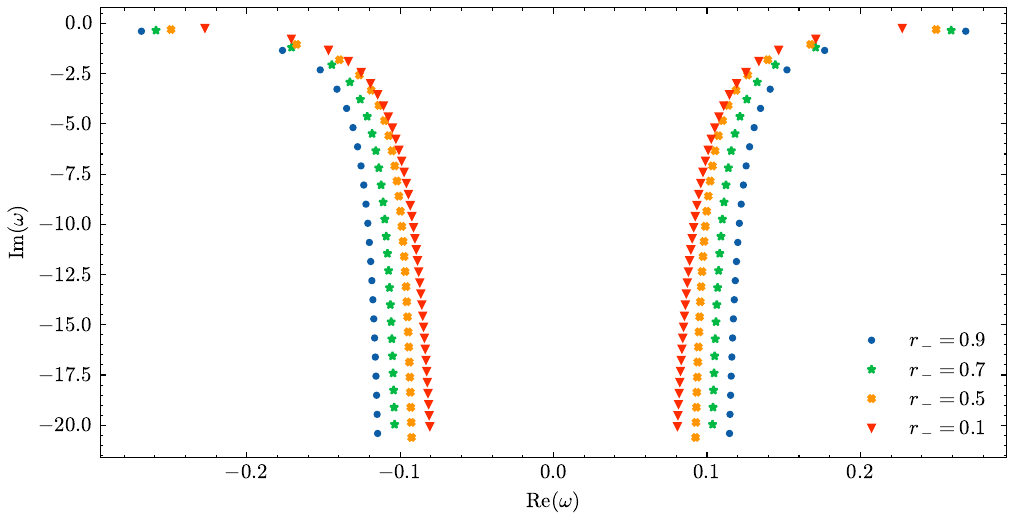}
        \caption{$r_-$ ranging from 0.1 to 0.9.}
    \end{subfigure}
    \caption{QNM spectra for spin $s=0$ test-field perturbations on the BCL BH for $l=0$. }
    \label{QNM_s0}

\bigskip

\begin{subfigure}{0.43\textwidth}
        \centering
        \includegraphics[width=0.85\linewidth]{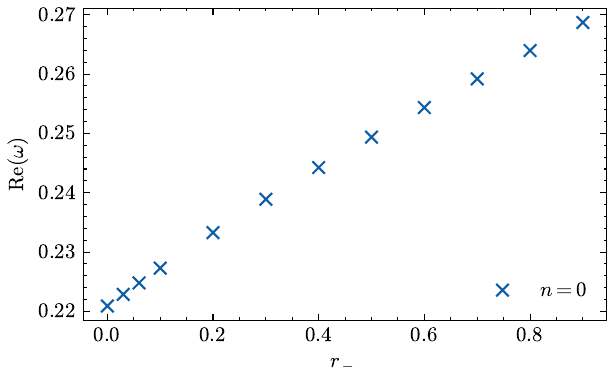}
    \end{subfigure}
    \begin{subfigure}{0.43\textwidth}
        \centering
        \includegraphics[width=0.85\linewidth]{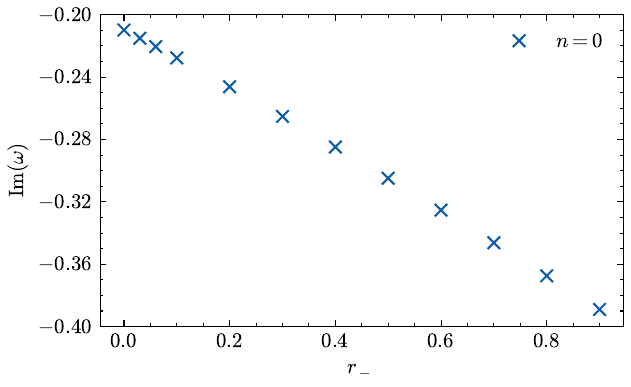}
    \end{subfigure}
    \\
    \begin{subfigure}{0.43\textwidth}
        \centering
        \includegraphics[width=0.85\linewidth]{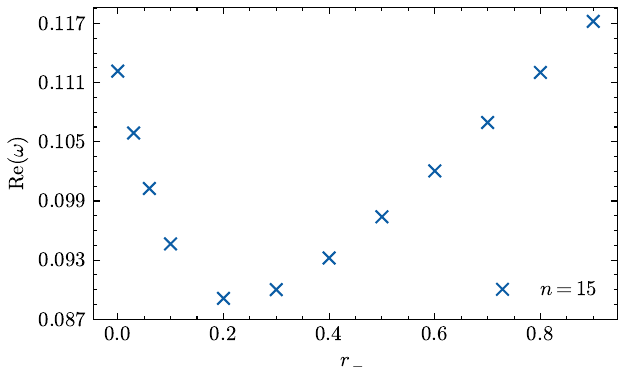}
    \end{subfigure}
    \begin{subfigure}{0.43\textwidth}
        \centering
        \includegraphics[width=0.85\linewidth]{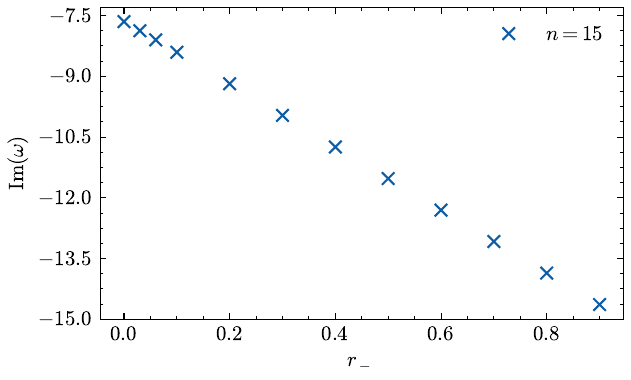}
    \end{subfigure}
    
    \caption{Evolution of the Real (\textit{left}) and Imaginary (\textit{right}) parts of the scalar (spin $s=0$) QNMs with $r_-$, for the fundamental mode (\textit{up}) and the 15th overtone (\textit{down}).}
    \label{QNM_s0_Re_Im}

\end{figure}
\begin{figure}[htbp]
\centering

\begin{subfigure}{0.73\textwidth}
        \centering
        \includegraphics[width=.85\linewidth]{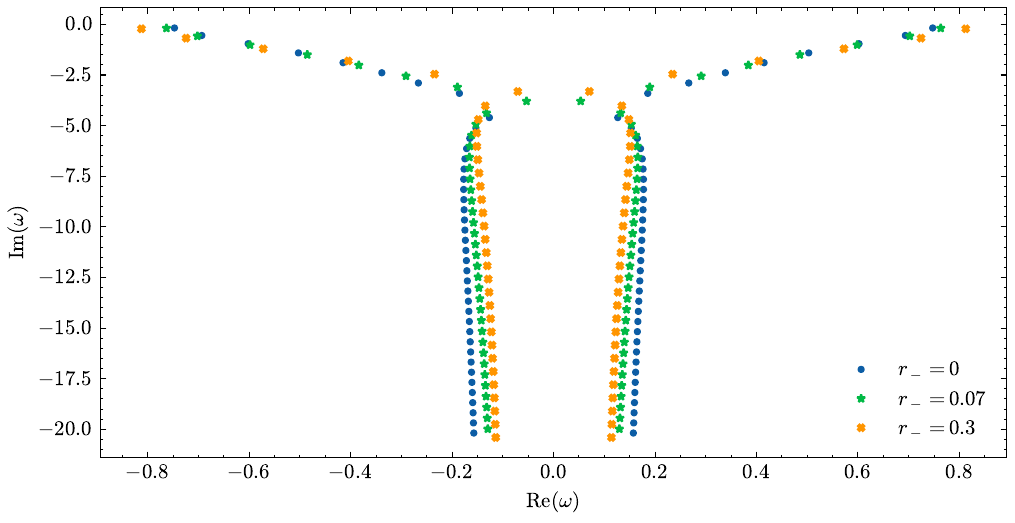}
        \caption{$r_-$ ranging from 0 to 0.3.}
    \end{subfigure}\\
    \begin{subfigure}{0.73\textwidth}
        \centering
        \includegraphics[width=.85\linewidth]{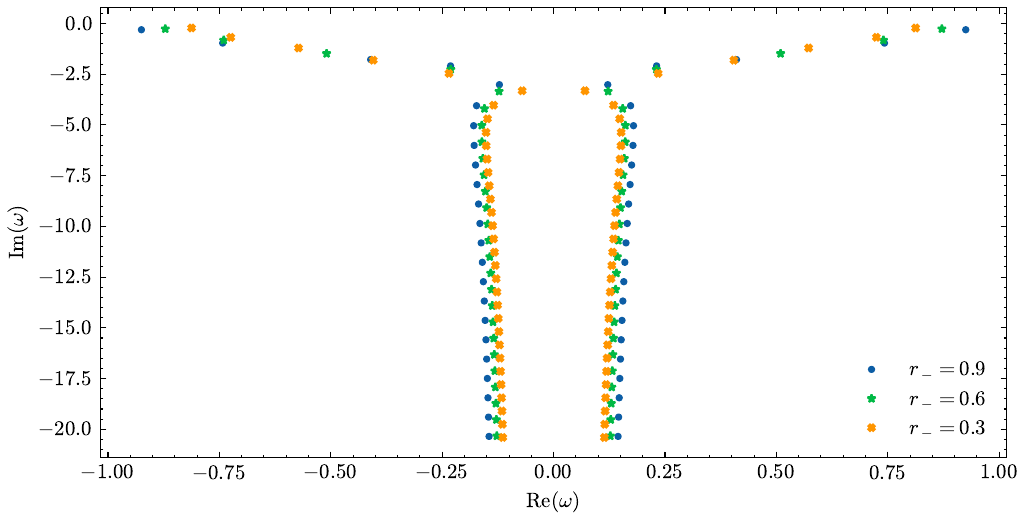}
        \caption{$r_-$ ranging from 0.3 to 0.9.}
    \end{subfigure}
    \caption{QNM spectra for spin $s=2$ test-field perturbations on the BCL BH for $l=2$.}
    \label{QNM_s2}

\bigskip

\begin{subfigure}{0.43\textwidth}
        \centering
        \includegraphics[width=0.85\linewidth]{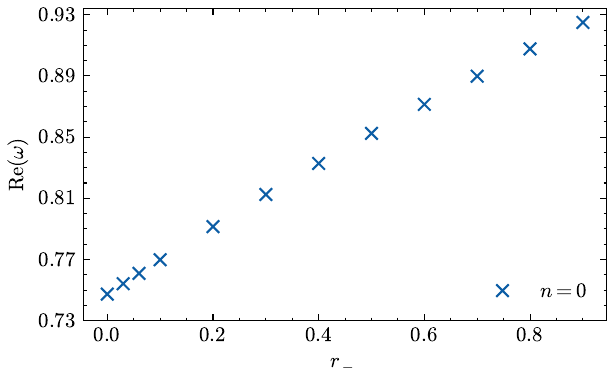}
    \end{subfigure}
    \begin{subfigure}{0.43\textwidth}
        \centering
        \includegraphics[width=0.85\linewidth]{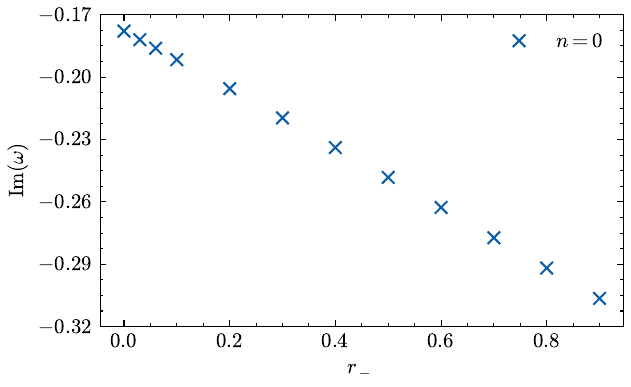}
    \end{subfigure}
    \\
    \begin{subfigure}{0.43\textwidth}
        \centering
        \includegraphics[width=0.85\linewidth]{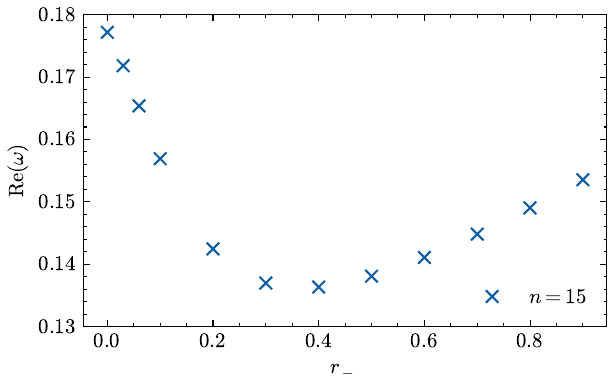}
    \end{subfigure}
    \begin{subfigure}{0.43\textwidth}
        \centering
        \includegraphics[width=0.85\linewidth]{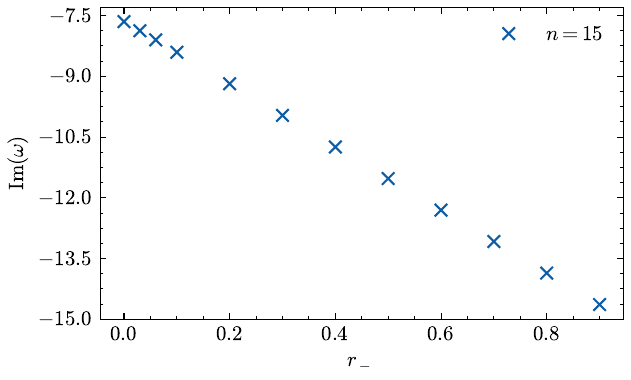}
    \end{subfigure}
    
    \caption{Evolution of the Real (\textit{left}) and Imaginary (\textit{right}) parts of the spin $s=2$ test-field QNMs with $r_-$, for the fundamental mode (\textit{up}) and the 15th overtone (\textit{down}).}
    \label{QNM_s2_Re_Im}

\end{figure}

In this subsection we will present QNM spectra highlighting the main properties of test-field QNMs for the BCL BH. QNM spectra for spin $s=0$ test-field perturbations are shown in figure \ref{QNM_s0} for $l=0$, while QNM spectra for spin $s=2$ are displayed in figure \ref{QNM_s2} for $l=2$. For each spin, the results are then split in two different spectra with different ranges of $r_-$ in order to have a better vision of what is happening. \\
To begin with, the fundamental mode has a relatively simple behaviour. Starting from the Schwarzschild frequency ($r_-=0$), the increase of the value of the BCL parameter $r_-$ linearly raises the absolute real part of the fundamental QNM and slightly decreases the imaginary part. \\
Then the behaviour of the higher modes becomes more intricate and depends on the value of $r_-$. Indeed for small values of $r_-$ ($r_- \lesssim 0.2$), the absolute real part start decreasing linearly with $r_-$ from the first overtone while it keeps increasing with $r_-$ for higher values of $r_-$. This results in \enquote{bouncing} spectra as one can see on figures \ref{QNM_s0} and \ref{QNM_s2}. This phenomena could be linked with the discontinuity observed in the monodromy results: the BCL asymptote result \eqref{mono_BCL} (in $\log(2)$) does not reduce to the Schwarzschild result \eqref{Sch_mono} in $\log(3)$.
\\
Note that despite this bouncing phenomena, the spectra for spin $s=0$ and $s=2$ test-field BCL QNMs keep the same general shape as the Schwarzschild spin $s=0$ and $s=2$ QNM spectra.

\begin{figure}[!ht]
    \centering
    \begin{subfigure}{0.45\textwidth}
        \centering
        \includegraphics[width=\linewidth]{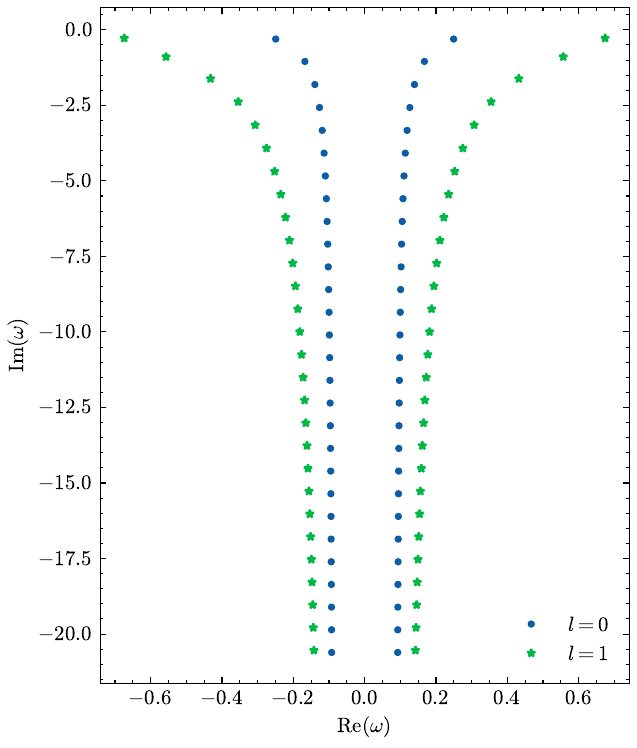}
        \caption{spin $s=0$, $l=0$ \textit{vs} $l=1$.}
    \end{subfigure}
    \begin{subfigure}{0.45\textwidth}
        \centering
        \includegraphics[width=\linewidth]{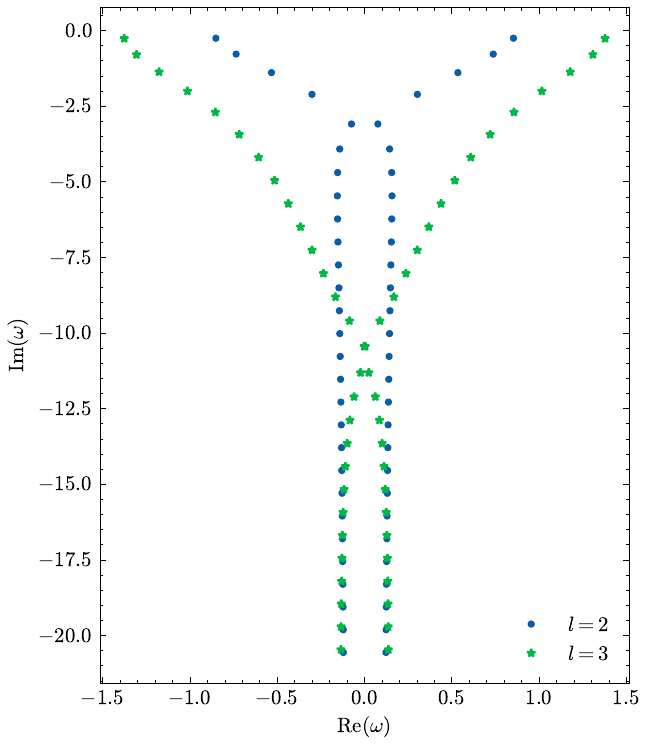}
        \caption{spin $s=2$, $l=2$ \textit{vs} $l=3$}
    \end{subfigure}
    \caption{QNM spectra for spin $s=0$ and $s=2$ test-field perturbations on the BCL BH for different values of the angular momentum parameter $l$.}
    \label{QNM_l}
\end{figure}

Now that we explored the effects of the BCL parameter $r_-$ on the test-field QNM spectra, let us look at the effect of the angular momentum $l$. One can see on figure \ref{QNM_l} that increasing $l$ has the same effect as for the Schwarzschild QNMs: the absolute real part of the first modes is increased by $l$, but both spectra seem to tend towards the same asymptotic. This is consistent with our analytical results for the asymptotic QNMs as we will see in the following section.

\subsection{Asymptotic spectrum: analytical results \textit{vs} numerical results}
\label{Comp_Asymp}

In this section we want to test the formula \eqref{mono_BCL} that we derived via the monodromy technique for the asymptotic QNMs of the BCL BH. The equivalent equation for the Schwarzschild BH \eqref{Sch_mono} was derived by Motl in 2002 \cite{Motl:2002hd}, almost ten years after it was numerically predicted by Nollert \cite{Nollert:1993zz}. Following Leaver who managed to compute 60 Schwarzschild QNMs, Nollert adapted the Continued Fraction method to the asymptotic part of the spectrum and succeeded to compute 2000 QNMs. This large amount of frequencies enabled him to correctly predict the first asymptotic order \eqref{Sch_mono} and to identify a second order term in $n^{-1/2}$ ($n$ being the mode number), depending on the angular momentum parameter $l$.

In the case of the test-field perturbations of the BCL, we have been unable to compute more than 30 to 100 QNMs, depending on the value of the parameter $r_-$. Despite our efforts to adapt the Leaver method to this case, we faced several issues which prevented us from finding more highly-damped QNMs. The main matter is the difference in the order of magnitude which appears between the real part Re$(\om)\sim 0.1$ and the imaginary part Im$(\om)\sim n$, as the QNMs become more and more damped. This seems to create issues in the process of Gauss reduction and computation of the roots of the continued fraction equation and its inversions. For example, we end up finding asymptotic QNMs with a consistent gap between the imaginary parts, but with a constant real part. It also happens to find unstable branches of QNMs, with different real and imaginary parts. The last case, which is fairly common, is that the computation time is far too long to be acceptable. The number of QNMs we were able to compute for each value of $r_-$ is giving in table \ref{table_nb_qnm} for indication. 

\begin{figure}[h!]
    \centering
    \begin{tabular}{|c||c|c|c|c|c|c|c|c|c|c|c|c|c|c|c||}
    \hline 
    $r_-$ &0&0.02&0.04&0.06&0.08&0.1&0.2&0.3&0.4&0.5&0.6&0.7&0.8&0.9\\
    \hline
    \hline
    $s=0$ & 200&  96&  95&  92&  85&  74&  51&  40&  36&  33&  34&  35&  41&
        65\\
    \hline
    $s=2$ &  200& 104& 105&  68&  96&  80&  51&  40&  33&  33&  32&  34&  42&
        56\\
    \hline
    \end{tabular}
    \caption{Number of spin $s=0,2$ test-field QNMs we were able to compute using the Continued Fraction method for each value of the BCL parameter $r_-$. The number of QNMs for the Schwarzschild BH ($r_-=0$) was chosen to stay in the same order of magnitude than for the BCL BH.}
    \label{table_nb_qnm}
\end{figure}

\begin{figure}[h!]
    \centering
    \begin{subfigure}{\textwidth}
        \centering
        \includegraphics[width=0.45\linewidth]{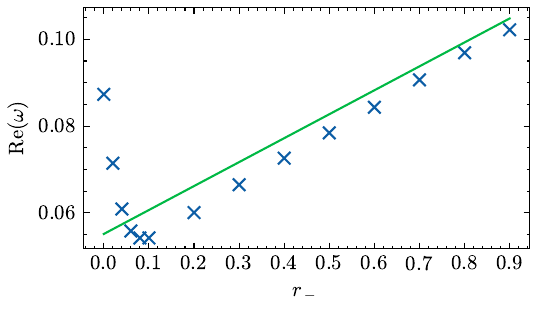}
        \hspace*{5mm}
        \includegraphics[width=0.45\linewidth]{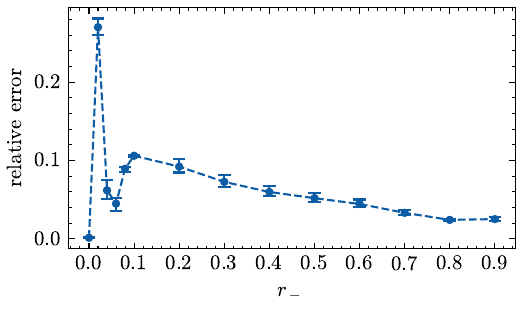}
        \caption{Comparison of the fitted values for the Real part at asymptotic with the monodromy prediction.}
    \label{Re_s0_asymplot}
    \end{subfigure}
    \begin{subfigure}{\textwidth}
        \centering
        \includegraphics[width=0.45\linewidth]{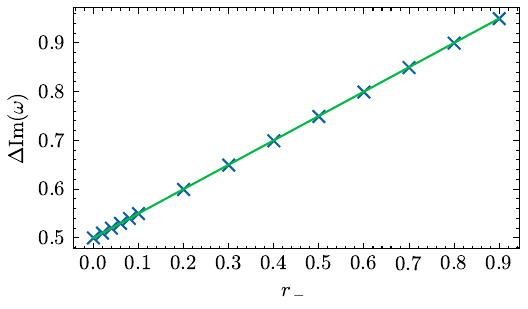}
        \hspace*{5mm}
        \includegraphics[width=0.45\linewidth]{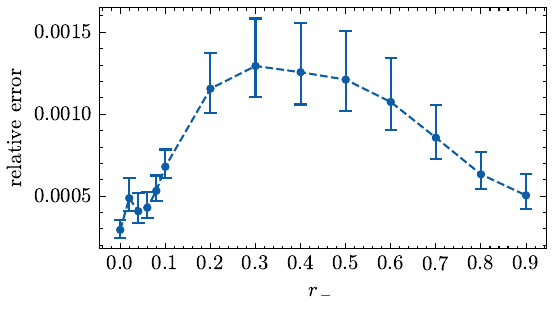}
        \caption{Comparison of the fitted values for the gap between two successive Imaginary parts at asymptotic with the monodromy prediction.}
    \label{Im_s0_asymplot}
    \end{subfigure}
    \caption{Evolution of asymptotic scalar QNMs with $r_-$. \textit{Left}: The blue crosses correspond to the constant value fitted from the asymptotic part of the QNM spectra obtained via the Leaver method (the first 20 QNMs were removed for this asymptotic analysis). The green line is the monodromy prediction for $r_->0$.\textit{Right}: Relative error between the fitted values and the monodromy prediction. Note that the relative error for $r_-=0$ is computed with the monodromy prediction for the Schwarzschild BH. The error bars were obtained by varying the number of low damped QNMs which were removed for the asymptotic analysis (between 15 and 25).}

    \label{asymp_plots_s0}
\end{figure}

\begin{figure}[!ht]
    \centering
    \begin{subfigure}{\textwidth}
        \centering
        \includegraphics[width=0.45\linewidth]{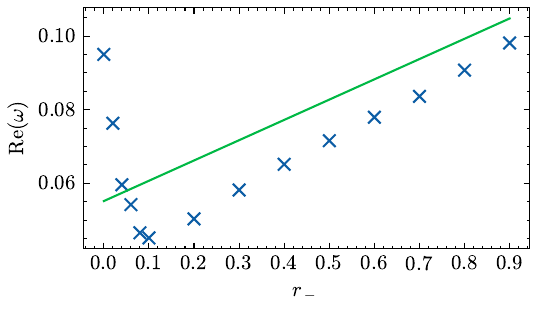}
        \hspace*{5mm}
        \includegraphics[width=0.45\linewidth]{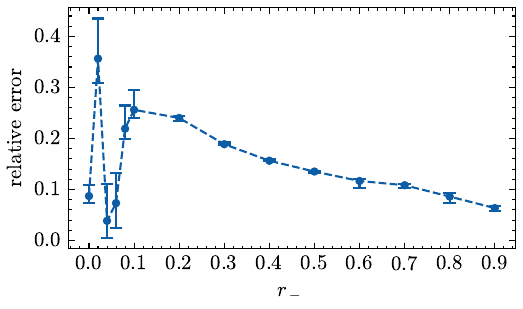}
        \caption{Comparison of the fitted values for the Real part at asymptotic with the monodromy prediction.}
    \label{Re_s2_asymplot}
    \end{subfigure}
    \\
    \begin{subfigure}{\textwidth}
        \centering
        \includegraphics[width=0.45\linewidth]{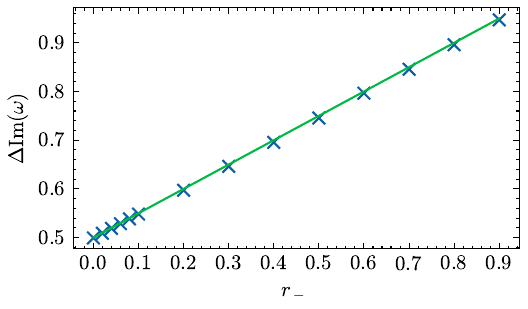}
        \hspace*{5mm}
        \includegraphics[width=0.45\linewidth]{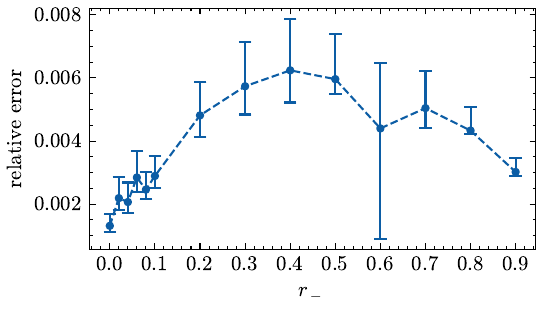}
        \caption{Comparison of the fitted values for the gap between two successive Imaginary parts at asymptotic with the monodromy prediction.}
    \label{Im_s2_asymplot}
    \end{subfigure}
    \caption{Evolution of asymptotic $s=2$ test-field QNMs with $r_-$. \textit{Left}: The blue crosses correspond to the constant value fitted from the asymptotic part of the QNM spectra obtained via the Leaver method (the first 20 QNMs were removed for this asymptotic analysis). The green line is the monodromy prediction for $r_->0$. \textit{Right}: Relative error between the fitted values and the monodromy prediction. Note that the relative error for $r_-=0$ is computed with the monodromy prediction for the Schwarzschild BH. The error bars were obtained by varying the number of low damped QNMs which were removed for the asymptotic analysis (between 15 and 25).}
    \label{asymp_plots_s2}
\end{figure}

One can note that the number of QNMs computed decreases as $r_-$ increases (setting apart the last two values 0.8 and 0.9 for which there is a small boost). This can partly be explained by the increase of the imaginary part gap with $r_-$ -- as one can see on figures \ref{asymp_plots_s0} and \ref{asymp_plots_s2} -- since it means that, for the same number of QNMs, the spectrum is spread out further down.

In order to compare the QNM spectra we obtained via the Continued Fraction method with the results of the monodromy technique for the BCL BH, we chose to analyse separately the Real and Imaginary parts of the QNMs at the asymptotic part of the spectra. Given the shape of the spectra, it is more suitable to study the \enquote{gap} between the Imaginary parts of two successive QNMs rather than the Imaginary part itself. We define this gap $\Delta$Im as:
\be 
\Delta \text{Im} \equiv \text{Im}(\om_n-\om_{n+1}).
\ee
The monodromy formula \eqref{mono_BCL} predicts that, at first order, $\Delta$Im$=\frac{1}{2R}$, where $R=\frac{r_+^2}{r_++r_-}$. When the parameter $r_-$ vanishes, we recover the famous Schwarzschild value $\Delta$Im$^{(\text{Sch})}=\frac{1}{2r_s}$.\\
In the case of the Real part, there is no continuity between the value given by the BCL monodromy formula \eqref{mono_BCL} for $r_-\rightarrow0$ and the Schwarzschild monodromy value \eqref{Sch_mono}. As already stated, this issue is due to the fact that we restrained to the first order of the BCL effective potential for the monodromy computation, and that this term vanishes when we take the Schwarzschild limit $r_-\rightarrow0$. It will then be interesting to see how the Real part of the asymptotic QNMs behaves in comparison with the monodromy predictions, especially for the small values of $r_-$.


In order to study the asymptotic part of the spectra, we removed the first 15 to 25 QNMs of the spectra for each value of $r_-$ and each spin $s=0,2$. We did not remove more than 25 modes as there are barely more than 30 QNMs available for some values of $r_-$ (see table \ref{table_nb_qnm}). In addition, 15 modes seemed to be the minimum acceptable in order to focus on the asymptotic \enquote{vertical} part of the spectra and ensure a good analysis.
\\
We then performed a curve fitting on the Real part and the Imaginary gap of each of these QNMs data collections, using the following ansatz function:
\be 
f(n) = A+\frac{B}{\sqrt{n}},
\label{fit}
\ee 
where $n$ denotes the QNM number. \\
The inverted square-root term was first motivated by the second order result found for the Schwarzschild BH (infered numerically by Nollert \cite{Nollert:1993zz} and later proved analytically by Musiri and Siopsis via the monodromy technique \cite{Musiri:2003bv}). It turned out to be the best fit according to our tests. We then applied a curve fitting with the function \eqref{fit} to the Real parts and the Imaginary gaps of the asymptotic data for spins $s=0,2$. The limited data we have did not allow us to infer the coefficient $B$, so we only focused on the constant one $A$.
\\
We present our results in figure \ref{asymp_plots_s0} for the scalar QNMs and in figure \ref{asymp_plots_s2} for the spin 2 test-field QNMs, with two different types of plots. 
Firstly, the left plots directly show the curve fitting results. The blue arrows represent the values of the constant $A$ \eqref{fit} found by curve fitting for each value of the parameter $r_-$ and by removing the first 20 modes of each spectra. A green line also appears: it represents the evolution of the first order asymptotic value \eqref{mono_BCL} predicted by the monodromy technique for the BCL BH (\textit{i.e.} strictly for $r_->0$).
Secondly, the plots on the right side show the evolution of the relative errors between the values obtained from the curve fitting of the QNM spectra and the values predicted by the monodromy computation, either for the Real parts or for the Imaginary gap. We added error bars corresponding to the values obtained by removing either 15 or 25 low-damped modes from the spectra, instead of 20.
Note that the relative errors for the Schwarzschild case ($r_-=0$) were computed using the monodromy prediction found for the Schwarzschild BH \eqref{Sch_mono} and not the one for the BCL BH. One can see that the results for the Schwarzschild BH are very good overall. In the case of scalar QNMs, the relative errors for the Real part and the Imaginary gap are around $0.1\%$ and $0.02\%$ respectively. The results for the spin $s=2$ test-field QNMs are slightly less good: about $8\%$ and 0.1$\%$ respectively. This difference with the scalar case can be explained with the shape of the spectra: the spin 2 spectra shows a crossing, as one can see in figure \ref{QNM_s2}, and in consequence the asymptote comes slightly later than in the scalar spectra. 
This also shows that 200 modes are not enough to achieve a very good match with the monodromy prediction (as Nollert successfully did with 2000 modes \cite{Nollert:1999ji}). As the data for the test-field perturbations of the BCL BH ranges between 30 and 100 modes, we expect higher errors values than the ones we obtained for the Schwarzschild case. 

Let us now turn to the analysis of our results for the case $r_->0$ and let us start with the Real part case. As we highlighted before, there is a big gap between the Schwarzschild value predicted by the monodromy technique ($\frac{\log(3)}{4\pi}$) and the one we found for the BCL BH ($\frac{\log(2)}{4\pi}(1+r_-)$). 
This difference is particularly noticeable on figures \ref{Re_s0_asymplot} and \ref{Re_s2_asymplot}, especially for the point corresponding to $r_-=0.02$ (for which the relative error is about $27\%$ for spin 0 and $35\%$ for spin 2). Then one can see that the Real asymptotic values experience a transitional regime when $r_-$ is increasing: the fitted value drops below the monodromy value around $r_-\sim0.1$ before getting closer and closer as $r_-$ tends to 1. In consequence, the relative error becomes artificially low for $0.4\lesssim r_-\lesssim0.6$ (around $5\%$) before increasing again (up to $10\%$ and $25\%$ respectively) and slowly lowering down until reaching $2\%$  to $6\%$ respectively for $r_-=0.9$.

For the case of the Imaginary gap, one can see that the correspondences between the numerical results and the monodromy prediction are much better than for the Real part. This can be explained by the fact that the first order for the Imaginary part scales in $n$ and that there is no discontinuity between the Schwarzschild formula \eqref{Sch_mono} and the one for the BCL BH \eqref{mono_BCL}. The relative errors start from 0.02$\%$ for the scalar case and $0.1\%$ for the spin 2 case, and reach at most $0.13\%$ and $0.6\%$ respectively. Those results are very satisfying and show that fifty to a hundred of modes are enough to infer the first order
asymptotic value of the Imaginary part.
\\

In conclusion, we can say that the numerical results are globally in good agreement with the monodromy predictions, even though the relative errors for the Real part go up to 40$\%$. Many more asymptotic modes would be required to obtained better results. However, as we highlighted earlier, there is a discontinuity between the monodromy formulas of the Schwarzschild and BCL BHs and the numerical analysis allowed us to visualize the transitory behavior of the asymptotic QNMs from one value to the other. In addition, the \enquote{bouncing} of the real part with $r_-$ is very similar to the one we observed in figures \ref{QNM_s0_Re_Im} and \ref{QNM_s2_Re_Im}, showing the continuity of this phenomenon in the complex spectrum. What would be interesting to do next is to compute the asymptotic second order with the monodromy technique, in order to better understand the transitionary \enquote{bouncing} behavior.

\section{Spin-2 test-field perturbations \textit{vs} physical gravitational perturbations}

\subsection{Axial perturbations of the BCL BH}

In this section, we summarize the computation process of the equations of motion for the linear perturbations of the BCL BH and the derivation detailed in (\cite{Langlois:2021aji,Roussille:2023sdr}), along with a short review of the matrix continued fraction method used in \cite{Roussille:2023sdr} to compute the physical QNMs.

In order to derive the equations of motion for the linear perturbations of the BCL BH, one must substitute in the BCL action \eqref{BCL_action} the perturbed metric along with the perturbed scalar field 
\be
g_{\mu \nu}=\bar{g}_{\mu \nu}+h_{\mu \nu}, \quad \phi=\bar{\phi}+\delta \phi,
\ee
where $\bar{g}_{\mu \nu}$ and $\bar{\phi}$ are respectively the metric and scalar field of the background, while $h_{\mu \nu}$ and $\delta \phi$ are the corresponding perturbations.

One then need to expand the action up to quadratic order in $h_{\mu \nu}$ and $\delta \phi$ in order to obtain the quadratic action governing the dynamics of the linear perturbations
\be
S_{\text {quad }}\left[h_{\mu \nu}, \delta \phi\right].
\ee
Finally, the Euler-Lagrange equations give the equations of motion for the linear perturbations:
\be
\mathcal{E}_{\mu \nu}=\frac{\delta S_{\text {quad }}}{\delta h_{\mu \nu}}=0 \quad \text { and } \quad \mathcal{E}_{\phi}=\frac{\delta S_{\text {quad }}}{\delta(\delta \phi)}=0.
\ee
The equation $\mathcal{E}_{\phi}=0$ is redundant due to Bianchi's identities.

In the following, only the perturbations for which the perturbed scalar field $\delta \phi$ is vanishing, \textit{i.e.} axial perturbations, are considered. As shown in \cite{Langlois:2021aji}, a first-order system is obtained in the following matrix form:
\be
\frac{\mathrm{d} X_{\mathrm{ax}}}{\mathrm{~d} r}=M_{\mathrm{ax}} X_{\mathrm{ax}},
\label{matrix_eq}
\ee 
with
\be 
X_{\mathrm{ax}}=\binom{h_{0}}{h_{1}}, \quad \quad M_{\mathrm{ax}}=\left(\begin{array}{cc}
\frac{2}{r} & -i \omega+\frac{2 i \lambda\left(r-r_{+}\right)\left(r+r_{-}\right)}{r^{4} \omega}  \\
-\frac{i \omega r^{2}\left(r^{2}+2 r_{-} r_{+}\right)}{\left(r-r_{+}\right)^{2}\left(r-r_{-}\right)^{2}} & -\frac{r\left(r_{+}-r_{-}\right)+2 r_{+} r_{-}}{r\left(r-r_{+}\right)\left(r+r_{-}\right)}
\end{array}\right).
\ee
The Schwarzschild limit is obtained for $r_{-}=0$.

Then the authors of \cite{Roussille:2023sdr} used what they called the matrix continued fraction method, an adaptation of the continued fraction method developed by Leaver \cite{Leaver:1985ax} for the matrix system \eqref{matrix_eq}.\\
The process is the same, starting with finding an ansatz satisfying the asymptotic behaviours of axial perturbations at both the horizon and spatial infinity. Those were computed in \cite{Langlois:2021aji}. The leading-order terms in the asymptotic expansion at spatial infinity (when $r \rightarrow \infty$ ) is given by
\begin{subequations}
\begin{align}
& h_{0}(r)=r\left(c_{+}^{\infty} e^{i \omega r} r^{i \mu \omega}-c_{-}^{\infty} e^{-i \omega r} r^{-i \mu \omega}\right)(1+\mathcal{O}(1 / r)), \\
& h_{1}(r)=r\left(c_{+}^{\infty} e^{i \omega r} r^{i \mu \omega}+c_{-}^{\infty} e^{-i \omega r} r^{-i \mu \omega}\right)(1+\mathcal{O}(1 / r)),
\end{align}
\end{subequations}
where $c_{ \pm}^{\infty}$ are constant.\\
Moreover, the asymptotic expansions near the horizon (when $r \rightarrow r_{+}$) yield, at leading order:
\begin{subequations}
\begin{align}
& h_{0}(r)=\left(c_{+}^{\mathrm{hor}}\left(r-r_{+}\right)^{+i \omega r_{0}}+c_{-}^{\mathrm{hor}}\left(r-r_{+}\right)^{-i \omega r_{0}}\right)\left(1+\mathcal{O}\left(r-r_{+}\right)\right), \\
& h_{1}(r)=\frac{r_{0}}{r_{+}}\left(-c_{+}^{\mathrm{hor}}\left(r-r_{+}\right)^{+i \omega r_{0}-1}+c_{-}^{\mathrm{hor}}\left(r-r_{+}\right)^{-i \omega r_{0}-1}\right)\left(1+\mathcal{O}\left(r-r_{+}\right)\right),
\end{align}
\end{subequations}
where $c_{ \pm}^{\text {hor }}$ are constant and $r_{0}$, which has the dimension of a radius, is defined by
\begin{equation}
r_{0}=r_{+} \frac{\sqrt{r_{+}\left(r_{+}+2 r_{-}\right)}}{r_{+}+r_{-}}.
\end{equation}
The QNMs boundary conditions (outgoing at infinity and ingoing at the horizon) lead to $c_{-}^{\infty}=0$ and $c_{+}^{\text {hor }}=0$ and the following ansatz is then appropriate for the description of axial perturbations:
\be
X_{\mathrm{ax}}(r)=e^{i \omega r} r^{i \mu \omega+1}\left(\frac{r-r_{+}}{r}\right)^{-i \omega r_{0}}\binom{f_{0}(u)}{f_{1}(u) / u}, \quad \text { with } \quad u=\frac{r-r_{+}}{r}.
\label{matrix_ansatz}
\ee
$f_{0}$ and $f_{1}$ are supposed to be bounded in the whole domain $u \in[0,1]$ and must satisfy the following boundary conditions
\be
f_{0}(0)=\frac{r_{+}}{r_{0}} f_{1}(0) \quad \text { and } \quad f_{0}(1)=-f_{1}(1).
\ee
Then, as it was done by Leaver \cite{Leaver:1985ax}, the two functions $f_{0}$ and $f_{1}$ are decomposed in power series,
\be
f_{0}(u)=\sum_{n=0}^{\infty} a_{n} u^{n} \quad \text { and } \quad f_{1}(u)=\sum_{n=0}^{\infty} b_{n} u^{n}.
\ee
Finally, injecting the ansatz \eqref{matrix_ansatz} into the matrix system \eqref{matrix_eq} gives a recurrence relation for the vector $Y_{n}$ defined as $Y_n=\binom{a_n}{b_n}$, involving 5 terms:
\be
\alpha_{n} Y_{n+1}+\beta_{n} Y_{n}+\gamma_{n} Y_{n-1}+\delta_{n} Y_{n-2}+\varepsilon_{n} Y_{n-3}=0, \quad n \geq 3 
\ee
The matrix coefficients along with more details for the QNMs computations can be found in \cite{Roussille:2023sdr}. As for the continued fraction method, Gaussian reduction is used to reduce the relation to a 3-term recurrence relation.

\smallskip

In the next section we will use the effective potential \eqref{V_P} given in \cite{Roussille:2023sdr}  of the Schrödinger-like equation \eqref{schro} derived from the first order matrix system \eqref{matrix_eq} in \cite{Langlois:2021aji}. This will allow us to compare it with the effective potential \eqref{V_F} of spin $s=2$ test-field perturbations and to use the WKB method for QNMs computations in appendix \ref{WKB_appendix}.

\subsection{Test-field effective potential \textit{vs} physical effective potential}

\begin{figure}[!ht]
     \centering
     \begin{subfigure}[b]{0.45\textwidth}
         \centering
         \includegraphics[width=.9\textwidth]{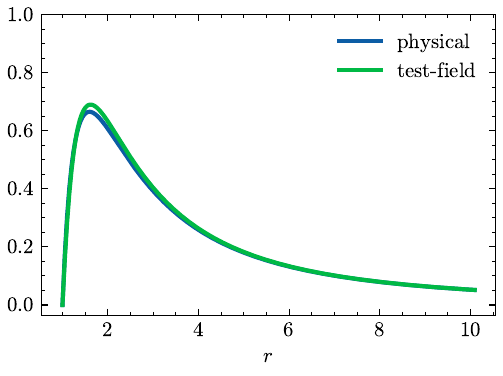}
         \caption{$r_-=0.2$}
     \end{subfigure}
     \hspace*{3mm}
     \begin{subfigure}[b]{0.45\textwidth}
         \centering
         \includegraphics[width=.9\textwidth]{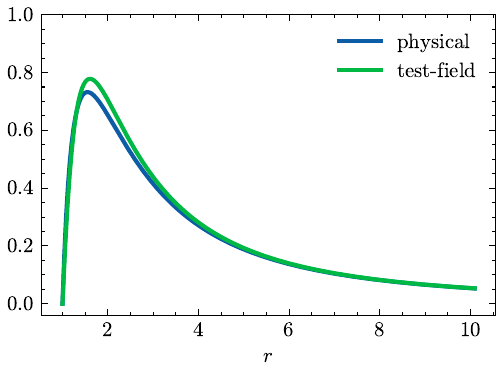}
         \caption{$r_-=0.4$}
     \end{subfigure}
     \hfill
     \begin{subfigure}[b]{0.45\textwidth}
         \centering
         \includegraphics[width=.9\textwidth]{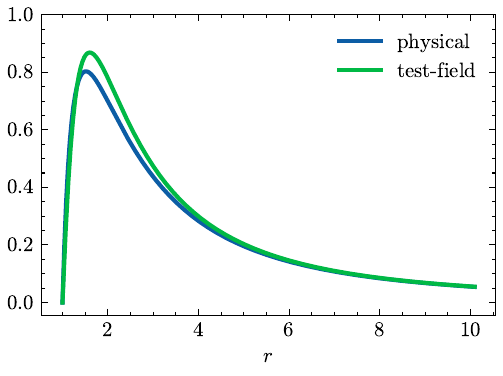}
         \caption{$r_-=0.6$}
     \end{subfigure}
     \hspace*{3mm}
     \begin{subfigure}[b]{0.45\textwidth}
         \centering
         \includegraphics[width=.9\textwidth]{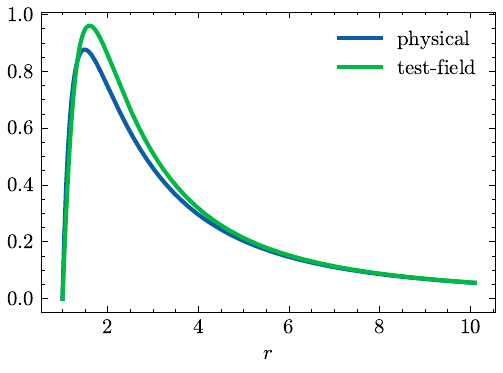}
         \caption{$r_-=0.8$}
     \end{subfigure}
     \caption{Comparison of the radial evolution of the effective potentials $V_{\text{phys}}$ and $V_{\text{field}}$ for different values of $r_-$.}
     \label{potentials}
\end{figure}

\ \\
Let us first recall the effective potential \eqref{BCL_pot} for a massless spin $s=2$ field in the BCL metric:
\be
V_{\text{field}}(r)=\frac{f(r)}{r}\lc\frac{\lambda}{r}-\frac{3(r_+-r_-)}{r^2}-4\frac{r_-r_+}{r^3}\rc.
\label{V_F}
\ee
On the other side, the effective potential of gravitational perturbations for the BCL metric was derived in \cite{Langlois:2021aji}. In the case of the BCL BH, the propagation speed of the gravitational perturbations differs from the usual value corresponding to the propagation speed of light. It is given by
\be 
c(r)=\frac{r}{\sqrt{r^2+2r_-r_+}}.
\ee
In order to compare the test-field potential with the physical one, we want to normalise the propagation speed to $c=1$. This can be done by a radial rescaling and by defining a new tortoise coordinate such that:
\be 
\dd \tilde{x} = \frac{1}{c(r)}\dd x,
\ee 
where $x$ is the usual tortoise coordinate define by $\dd x = \frac{1}{f(r)}\dd r$.
The effective potential for a normalized propagation speed $c=1$ reads \cite{Langlois:2021aji,Roussille:2023sdr}
\be 
\begin{aligned}
V_{\text{phys}}(r)=\frac{f(r)}{r\lp 1+2\frac{r_- r_+}{r^2}\rp^3}\lc \frac{\lambda }{r}-\frac{3(r_+-r_-)}{r^2}-\frac{r_- r_+(6\lambda-15)}{r^3}+\frac{4r_- r_+(r_--r_+)}{r^4}+\frac{r_-^2r_+^2(12\lambda-30)}{r^5}\right. \\ \left.+\frac{3r_-^2r_+^2(r_--r_+)}{r^6}+\frac{r_-^3r_+^3(8\lambda-21)}{r^7}\rc.
\label{V_P}
\end{aligned}
\ee

\begin{figure}[!ht]
    \centering
    \includegraphics[width=0.5\linewidth]{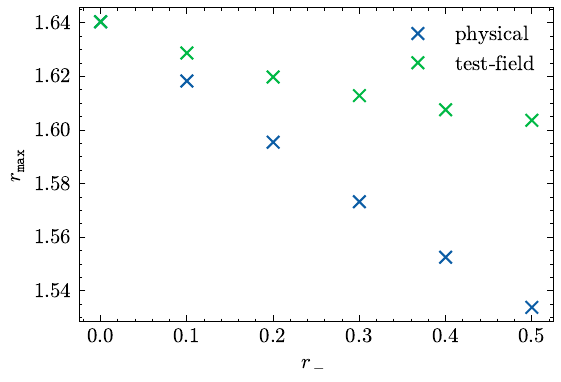}
    \caption{Comparison of the radial value $r_{\text{max}}$ for which the effective potential is maximum, for the both the test-field one and the physical one. $l$ is taken equal to 2.}
    \label{plot_r_max}
\end{figure}

The main difference with the potential $V_{\text{field}}$ is the presence of the factor $\lp 1 +2\frac{r_- r_+}{r^2} \rp^{-3}$, which reduces to 1 when $r$ is taken to infinity. Setting aside this, one can see that the terms of order 1 and 2 in $\frac{1}{r}$ are equal, while the term of order 3 exhibits a slight difference. New terms of order 4 to 7 appear in the potential $V_{\text{phys}}$ and show effects of the modified dynamic.
\\
Both physical and test-field potentials are shown in figure \ref{potentials} for $r_+=1$ and several values of $r_-$. Several comments can be made. First, the behaviour of both potentials at infinity seems to be very similar one to each other. This observation is supported by the analytical behaviour of each potential at infinity, which are almost equal up to the 3rd order:
\be
V_{\text{field}}\stackrel{\infty}{\sim} \frac{\lambda}{r^2}-(3+\lambda)\frac{r_+-r_-}{r^3}+\frac{3(r_-^2+r_+^2)-(\lambda+10)r_-r_+}{r^4}+\mathcal{O}\lp\frac{1}{r^5}\rp,
\ee
\be
V_{\text{phys}}\stackrel{\infty}{\sim} \frac{\lambda}{r^2}-(3+\lambda)\frac{r_+-r_-}{r^3}+\frac{3(r_-^2+r_+^2)-(\lambda+21)r_-r_+}{r^4}+\mathcal{O}\lp\frac{1}{r^5}\rp,
\ee
Second, the test-field and the physical potential show more differences when we look closer to the BH horizon. The most striking effect of the modified dynamics is the radial shift of the potential peak and its decrease in amplitude. One can see the detailed evolution of the potential peak $r_{\text{max}}$ in function of $r_-$ in figure \ref{plot_r_max}. In any case the peak location is decreasing with $r_-$, but the modified dynamics reduces this tendency.

Finally, one can see by zooming on figure \ref{potentials} right at the horizon that the slope of each potential is different. This can be related to the analytical behaviour at horizon of $V_{\text{field}}$ and $V_{\text{phys}}$:
\be 
V_{\text{field}}\stackrel{r_+}{\sim} \frac{(r-r_+)(r_-+r_+)}{r_+^5}\lc(\lambda-3)r_+-r_-\rc+\mathcal{O}\lp(r-r_+)^2\rp,
\ee
\be 
V_{\text{phys}}\stackrel{r_+}{\sim} \frac{(r-r_+)(r_-+r_+)}{r_+^4(2r_-+r_+)^2}\lc(\lambda-3)r_+^2 + (4\lambda-9)r_-^2+(4\lambda-10)r_-r_+\rc +\mathcal{O}\lp(r-r_+)^2\rp.
\ee
Indeed, unlike at infinity, both potential's expansions differ from the very first order. 
\\

The effective potentials for both physical and test-field perturbations exhibit important disparities, directly related to the modified dynamics. They most likely indicate variations to be observed in the QNM spectra. We will examine this in the next section, where we will compare the QNMs corresponding to the test-field effective potential $V_{\text{field}}$ to the ones computed via the physical potential $V_{\text{phys}}$ in \cite{Roussille:2023sdr}.

\subsection{Numerically computed QNMs for test-field perturbations \textit{vs} physical perturbations}

%
\begin{figure}[!ht]
     \centering
     \begin{subfigure}[b]{0.3\textwidth}
         \centering
         \includegraphics[width=\textwidth]{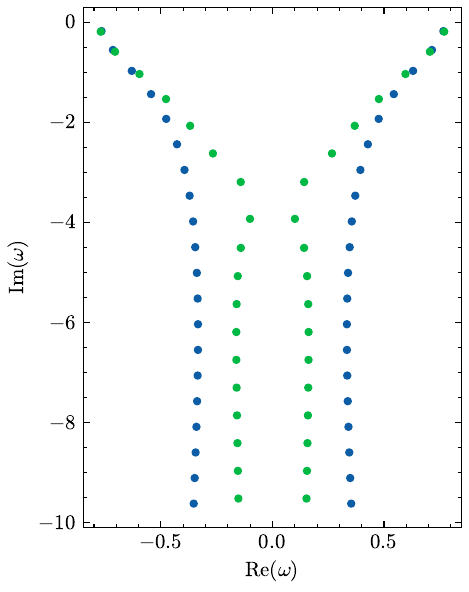}
         \caption{$r_-=0.1$}
     \end{subfigure}
     \begin{subfigure}[b]{0.3\textwidth}
         \centering
         \includegraphics[width=\textwidth]{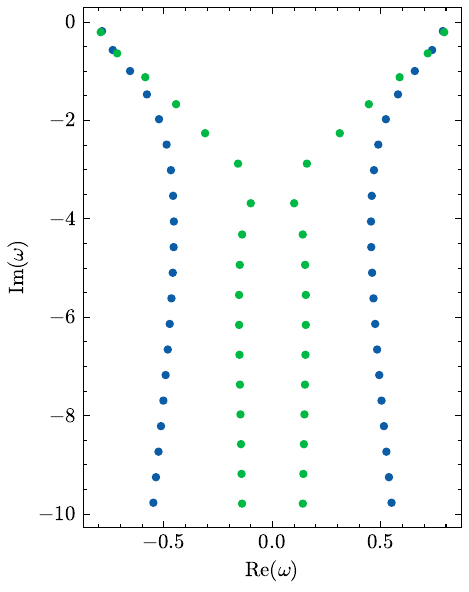}
         \caption{$r_-=0.2$}
     \end{subfigure}
     \begin{subfigure}[b]{0.3\textwidth}
         \centering
         \includegraphics[width=\textwidth]{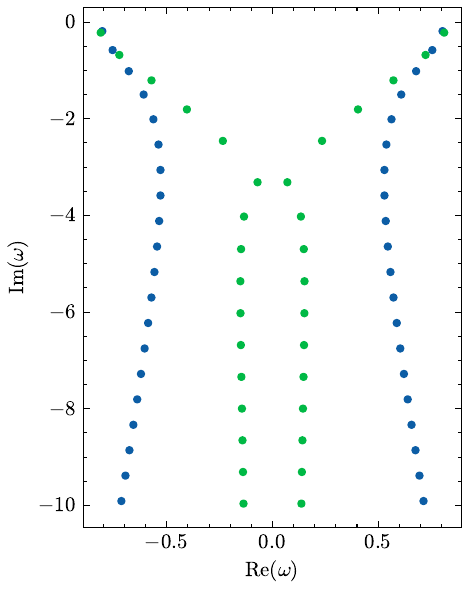}
         \caption{$r_-=0.3$}
     \end{subfigure}\\
     \begin{subfigure}[b]{0.3\textwidth}
         \centering
         \includegraphics[width=\textwidth]{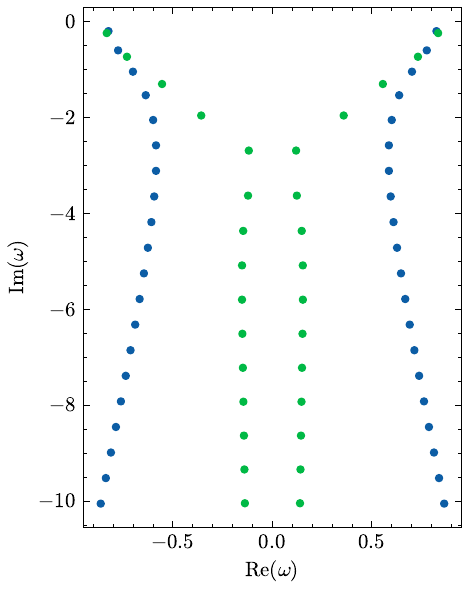}
         \caption{$r_-=0.4$}
     \end{subfigure}
     \begin{subfigure}[b]{0.3\textwidth}
         \centering
         \includegraphics[width=\textwidth]{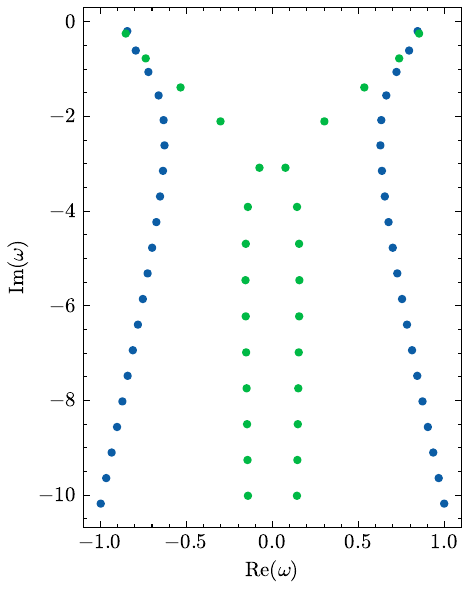}
         \caption{$r_-=0.5$}
     \end{subfigure}
     \caption{QNM spectra for physical perturbations in blue (\cite{Roussille:2023sdr}) and for $s=2$ test-field perturbations in green, for different values of $r_-$ and for $l=2$.}
     \label{plot_comp}
\end{figure}

\begin{figure}[!ht]
    \centering
    \includegraphics[width=0.5\linewidth]{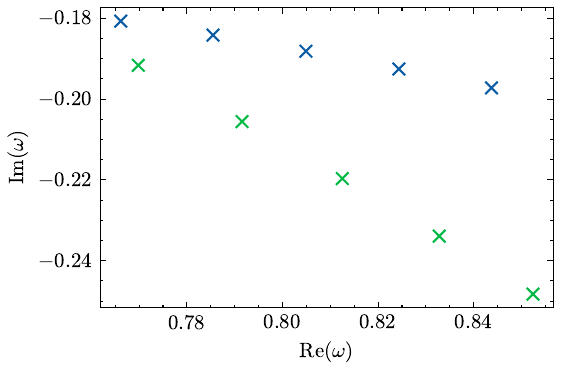}
    \caption{Fundamental QN mode of physical gravitational perturbations (in blue) \textit{vs} spin $s=2$ test-field perturbations (in green), for $r_-$ going from 0.1 to 0.5 (left to right) }
    \label{w0_comp}
\end{figure}

Let us now compare our numerical results for the spin $s=2$ test-field perturbations of the BCL with the numerical results obtained in \cite{Roussille:2023sdr} for physical gravitational perturbations of the same modified BH. An explicit comparison of the respective spectra is shown is figure \ref{plot_comp} for different values of $r_-$.\\
The first thing one can notice is the closeness of the values for the fundamental mode. To highlight this feature, we plotted the fundamental mode for each type of perturbations and for different values of $r_-$ in figure \ref{w0_comp}. One can see that for small value of $r_-$, both types of fundamental QNMs are very close to each other, while they distance themselves when $r_-$ is increasing. Typically, the test-field fundamental mode for $r_-=0.1$ is 0.78$\%$ different in absolute value from the physical one, while this figure goes up to 2.5$\%$ for $r_-=0.5$.
\\
Then, the following overtones deviate from each other more and more as we go down to larger imaginary parts, until displaying a completely different behaviour at the asymptote. For the spin $s=2$ test-field case, the QNM spectra are very similar in their shape to the Schwarzschild spectra, exhibiting a crossing of the imaginary axis and a vertical asymptote, the later being in fairly good agreement with the monodromy formula we derived. On the other side, the physical QNM spectra show no more crossing of the imaginary axis -- even for small values of $r_-$ -- and their asymptote becomes less and less vertical as $r_-$ increases. Roussille et al. parametrized the asymptote by the equation $\text{Im}(\omega) = a\times \text{Re}(\omega)+b$ \cite{Roussille:2023sdr} and plotted $1/a$ in function of $r_-$. From their data we evaluated the approximate dependence of $1/a$ in the parameter $r_-$: $1/a = -0.14(\pm0.01) r_- +0.04(\pm0.01) r_-^2$. However, this is only a numerical approximation. An analytical analysis of the asymptote would help characterizing it and better understanding the source of a non-vertical slope. Unfortunately, we were not able to perform a monodromy analysis for the physical perturbation case, as we did for the test-field case, due to the more complicated and subtler structure of the involved functions in the complex plane.

As long as the previous numerical analysis of the BCL BH \cite{Roussille:2023sdr} can be trusted, it turns out that the test-field spin 2 QNMs are substantially different from the physical gravitational ones, especially in the case of highly-damped modes. The case of the BCL BH is thus a great example of how test-field QNM spectra can differ from physical QNM spectra. It can be noted that, in this case, the test-field fundamental mode nevertheless constitute a very good approximation for the physical one. This can be linked to the fact that the behaviours of the effective potentials for each case are very similar at infinity. On the contrary, their expansion around the horizon $r\sim r_+$ is quite different, which could be the reason of the divergence in behaviour for the highly-damped QNMs.

\section*{\uppercase{Discussion and conclusion}}

In the context of the stability of Black Hole (BH) solutions in General Relativity (GR), the Quasi-Normal Modes (QNMs) made it possible to study in detail the response of BHs to any kind of perturbations and their return to an equilibrium state. Since the first detection of GWs ten years ago by LIGO \cite{LIGOScientific:2016aoc}, the QNMs regained a lot of interest and are now viewed as a direct way to study the physic of BHs and test new models.
Indeed, this advent of GW astronomy arrives in a context of intense search for a quantum theory of gravity and many different theories and models have flourished these last decades. The lack of direct test makes it difficult to discriminate those theories, but the QNMs are now allowing to characterise each theory and BH model with the hope that deviations from GR will be detected one day in the QNM spectrum detected via GWs.

Two main different directions are considered for the search of an alternative theory of gravity: modifying GR as in the scalar-tensor theories, or developing new microscopic space-time structures as in the String Theory and the LQG. Despite the lack of an equivalent of the Einstein equations for this last type of theories of gravity, the study of test-field perturbations on effective models of BHs via the QNMs has become a common practice. While this type of research allows to study the response by BHs to any type of test-field perturbations and compare it with GR, the link with physical gravitational perturbations and BH spectroscopy is not evident. 
With this question in mind, we wanted to compare test-field perturbations and physical gravitational ones for a \enquote{simple} BH in modified gravity for which the computation of these two types of perturbations is possible. 

Our choice came on the BCL BH \cite{Babichev:2017guv}, named from his authors. It is a solution of the quadratic shift-symmetric Horndeski theories, displaying a single additional parameter $r_-$ which describes the deformations from the Schwarzschild model. It can be considered as a simple universal template model for modified BHs with no difference in topology (i.e. no extra Cauchy surface, no new singularity and no singularity resolution).
QNMs of this BH model were already studied numerically by Roussille et al. \cite{Roussille:2023sdr}, focusing on gravitational perturbations derived directly from the modified Einstein equations. This gives an unique opportunity to compute test-field QNMs of the same BH, in order to compare them with these previous results and see to which extent they differ. 
We chose to focus on field perturbations of spin $s=0$ -- which are simply computed via the Klein Gordon equation and do not depend of the background theory of gravity -- and of spin $s=2$ to explicitly compare them to the physical gravitational perturbations ones.

For both cases, we performed several types of computations to determine QNMs with the best accuracy possible. First, we used the Continued Fraction method developed by Leaver \cite{Leaver:1985ax} in 1985. It is well-known for its ability to compute QNMs in a large part of the spectra. In the present paper, we were able to determine precisely low-damped and middle-damped QNMs for any value of $r_-$, but we had issues for the computation of highly damped QNMs. Typically, we computed QNMs down to -20i to -50i in the complex plane, the results being better for very low and very high values of $r_-$: $0 < r_- \lesssim 0.1$ and $0.9\lesssim r_- < 1$. We tried to adapt Leaver's method in various ways, but were unable to go further than the results presented here without enduring excessively long computation time. A significant obsctable was that the Gaussian reduction, necessary to handle the  recursion relations, makes it impossible to use the Nollert technique, and most likely leads to numerical errors when the Imaginary part of a QNM becomes very large compared to its Real part.

Nevertheless, our results allowed us to see multiple interesting behaviours. First, the spectra have the same kind of shape as the Schwarzschild one, whether for the spin 0 or the spin 2. Second, we noticed a sort of \enquote{bouncing} phenomena in $r_-$, starting from the middle-damped area of the spectra, that can be visualised in figures \ref{QNM_s0} to \ref{QNM_s2_Re_Im}. Comparing with the numerical data available in \cite{Roussille:2023sdr} for the physical QNMs, we noticed that the latter differ on these two points. In particular, while the low-damped modes are very similar for the two types of perturbations, the middle and high damped QNMs show a very different behaviour: the physical spectra show no crossing and no vertical asymptote in contrast to the test-field ones.

\smallskip

Moreover, we strengthened our analysis with other methods, both numerical and analytical.
On the one hand, we used the numerical WKB method to check the validity of our low-damped results.  Knowing that this technique is only accurate for $n \ll l$, we found out a good matching between the two techniques. The results are displayed in appendix \ref{WKB_appendix}. In their paper \cite{Roussille:2023sdr}, Roussille et al. also used the WKB method as a cross-check but only for $l=2$, for which the matching is not much precise. We then also applied the WKB method to the physical potential \eqref{V_P} to first, check the results of \cite{Roussille:2023sdr} for $l=2$ and then, compute the associated QNMs for higher values of $l$ (10 and 100). While the results between the WKB technique and the Continued Fraction method perfectly match for $V_{\text{field}}$ in the eikonal limit ($l=100$ here), the QNMs values obtained via the WKB technique applied on the physical potential $V_{\text{phys}}$ appear to be quite different from them. To be precise, one can see that the Imaginary part is the one showing the largest deviation, while the Real part remains pretty close. 

On the other hand, we used the monodromy technique to derive the asymptotic first order value of high-damped QNMs. We obtained a result similar in appearance to that of Schwarzschild, within a factor $R$ \eqref{mono_BCL}, except that we obtained a Log(2) in the Real part instead of a Log(3) as for Schwarzschild. There is then no continuity between both formulas. Yet, this does not mean our result is inconsistent, but rather that there exists some transitory behaviour which would appear if we would be able to go at the next order. The same type of discontinuity was observed with the first order monodromy result of the Reissner-Nordström \cite{Motl:2003cd}.
We have then been able to compare this theoretical prediction with our numerical results by performing a curve fitting on the \enquote{vertical} part of the QNM spectra. All the details about this fit have been given in the subsection \ref{Comp_Asymp}. The comparison between the monodromy and our numerical results turned very good for the Imaginary part and less for the Real part, mostly because of the transitory behaviour and the lack of high-damped modes. This comparison allowed us to better understand the phenomena of \enquote{bouncing} in $r_-$ that we already observed before. Nevertheless, there is much more to be understood about its mathematical origin (can it be seen directly from the shape of the effective potential \eqref{V_F} -- knowing in particular that this phenomena does not seem to happen for the physical QNMs?) and about its physical impact on modified gravity and BHs.

\bigskip

The aim of this paper was to perform a direct comparison between test-field perturbations and physical ones, and to visualize the differences on the QNM spectra in the case of the BCL BH from Modified Gravity. Thanks to the Continued Fraction method, the WKB approximation, the monodromy technique and the near-horizon $\sl_{2}$ symmetry, we have been able to compute spin 0 and 2 test-field QNMs with good accuracy and to highlight the differences we observed with the physical gravitational QNMs computed in \cite{Roussille:2023sdr}. The more damped the QNMs are, the more differences we can see. In particular, the asymptote line appears to be very different (non-vertical for the physical QNMs) and it would be very interesting to have a mathematical understanding of this -- like a monodromy computation as we did for the test-field case. 
In the BCL case, we noticed that the effective potentials were similar at infinity but quite different around the horizon, which could explain the differences that can be observed in the QNM spectra.

The natural question is, then, whether this feature is specific to this example or general to modified BH models. This means understanding precisely which BH properties the QNMs actually probe, i.e. if they are sensitive to the asymptotic region, or the near-horizon region, or the inner region (even if it is in principle inaccessible to exterior observers). More generally, this would lead to understanding for which classes of BH models test-field QNMs can be good approximations for physical perturbations, beyond our analysis of the \enquote{simple} example of the BCL metric.
%
This line of research appears crucial in the context of the multiplication of effective BH models and the booming development of BH spectroscopy.

\
\medskip
\medskip
\

\begin{center}
    {\textbf{Acknowledgments}}
\end{center}

We are grateful to Hugo Roussille and Karim Noui for explanations and discussions about the BCL black hole ansatz and the computation of quasi-normal modes.

\appendix

\section{Third order WKB approximation check}
\label{WKB_appendix}

The WKB method is well known in quantum mechanics where it is used to study solutions of the Schrödinger equation. It has been first applied to the problem of black holes scattering by Schutz and Will in 1985 \cite{Schutz:1985km}. This semi-analytic technique can be used for any barrier type effective potential having constant values at the boundaries. The idea is to match the WKB solution at the event horizon with the one at infinity with a Taylor series expansion. The third order approximation reads \cite{Iyer:1986nq}:
\be
\omega^2\approx V_0 + \sqrt{-2V_0''}\Lambda - i\lp n+\frac{1}{2}\rp \sqrt{-2V_0''}(1+\Omega),
\ee
where the coefficients $\Lambda$ and $\Omega$ are given in terms of the potential by
\beq
\Lambda &=&
\frac{1}{\sqrt{-2V_0''}}
\left[
\frac{1}{8}\lp \frac{V_0^{(4)}}{V_0''}\rp \lp \frac{1}{4}+\alpha^2\rp-\frac{1}{288}\lp\frac{V_0'''}{V_0''}\rp^2\lp 7+60\alpha^2)\rp
\right]
\,,\\
\Omega &=&
\frac{1}{-2V_0''}
\Bigg[
\frac{5}{6912}\lp\frac{V_0'''}{V_0''}\rp^4\lp 77+188\alpha^2\rp
-\frac{1}{384}\lp \frac{V_0'''^2 V_0^{(4)}}{V_0''^3}\rp \lp51+100\alpha^2\rp
\\
&&+\frac{1}{2304}\lp \frac{V_0^{(4)}}{V_0''}\rp^2\lp 67 +68 \alpha^2\rp +\frac{1}{288}\lp \frac{V_0'' V_0^{(5)}}{V_0''^2}\rp\lp 19+28\alpha^2\rp - \frac{1}{288}\lp\frac{V_0^{(6)}}{V_0''}\rp \lp 5+4\alpha^2\rp \Bigg]\,. \nn
\eeq
$\alpha$ is given by $\alpha = n+\frac{1}{2}$. The derivatives are taken with respect to the tortoise coordinate $x$ and the subscript $0$ indicates that the potential and its derivatives are evaluated at the maximum of the potential.\\
We use this technique to compute some QNM frequencies for $r_-=0.5$ and compare them to those computed thanks to the continued fraction method. 
First let us look at the the spin $s=0$ test-field QNMs: the results are displayed in table \ref{s0WKBcomp}. As expected, we see that the WKB approximation is not much accurate for small $l$ and $n\sim l$. Yet, the more $l$ is taken large, the better correspondence we obtain for the first QNMs.

\begin{table}[!ht]
    \centering
    \begin{tblr}{|c|c||c|c|}
    \hline
    $l$ & $n$ & WKB method for $V_{\text{field}}$ & Continued fraction for $V_{\text{field}}$\\
    \hline
    \hline
    0 & {0\\1} & {$0.230906 - 0.357451$i \\ $0.189902 - 1.07783$i} & {$0.249381 - 0.304853$i \\ $0.16755 - 1.0459$i} \\
    \hline
    1 & {0\\1\\2} & {$0.662001 - 0.278809$i \\ $0.539224 - 0.903158$i \\ $0.377397 - 1.56149$i}  & { $0.674058-0.276896$i \\ $0.556788-0.892044$i \\ $0.432179-1.61576$i} \\
    \hline
    2 & {0 \\1 \\ 2\\3} & { $1.11457 - 0.27356$i \\  $1.02886 - 0.845956$i \\ $0.896806 - 1.45273$i \\ $0.728963 - 2.07633$i} & { $1.11715 - 0.273238$i \\ $1.03392 - 0.843825$i \\ $0.897623 - 1.48062$i \\ $0.761482 - 2.19025$i} \\
    \hline
    10 & {0 \\ 1 \\  2\\3} & { $4.6791 - 0.271175$i \\ $4.65777 - 0.814913$i \\ $4.61606 - 1.3626$i \\ $4.55566 - 1.91621$i } & { $4.67913-0.271173$i \\ $4.65783-0.814903$i \\ $4.61556-1.36279$i \\ $4.55298-1.91763$i} \\
    \hline
    100 & {0 \\ 1 \\  2\\3} & { $44.7788 - 0.271047$i \\ $44.7766 - 0.813157$i \\ $44.7721 - 1.35531$i \\ $44.7654 - 1.89754$i } & { $44.7788-0.271047$i \\ $44.7766-0.813157$i \\ $44.7721-1.35531$i \\ $44.7654-1.89754$i} \\
    \hline
    \end{tblr}
    \caption{Spin $s=0$ test-field QNM frequencies for $r_-=0.5$, computed  using the third order WKB approximation and Leaver's method.}
    \label{s0WKBcomp}
\end{table}

\begin{table}[!ht]
    \centering
    \begin{tblr}{|c|c||c|c||c|c|}
    \hline
    $l$ & $n$ & WKB method for $V_{\text{field}}$  & Continued fraction for $V_{\text{field}}$ & WKB method for $V_{\text{phys}}$ &
    \begin{tabular}{@{}c@{}}Matrix continued fraction \\ (from \cite{Roussille:2023sdr})\end{tabular}
    \\
    \hline
    \hline
    2 & {0 \\1\\2\\3} & { $0.851619 - 0.251621$i \\ $0.740681 - 0.78549$i \\ $0.563749 - 1.35899$i \\ $0.329717 - 1.95672$i} & { $0.852397 - 0.248265$i \\  $0.736496 - 0.773467$i \\ $0.533955 - 1.38661$i \\ $0.301878 - 2.10415$i} & { $0.843204 - 0.198052$i \\ $0.793898 - 0.60744$i \\ $0.715865 - 1.03543$i \\ $0.615373 - 1.47301$i} & { $0.843719 - 0.197227$i \\ $0.794262 - 0.607533$i \\ $0.720858 - 1.059863$i \\ $0.661666 - 1.556088$i} \\
    \hline
    10 & {0 \\ 1 \\  2 \\3} & { $4.61896 - 0.269725$i \\ $4.59739 - 0.810575$i \\ $4.5552 - 1.35541$i \\ $4.49412 - 1.90621$i } & { $4.61899-0.269723$i \\ $4.59745-0.810564$i \\ $4.55469-1.3556$i \\ $4.49138-1.90767$i } & {$4.61909 - 0.221798$i \\ $4.60827 - 0.666182$i \\$4.58713 - 1.11282$i \\ $4.5566 - 1.5629$i} & { \text{no results available} }\\
    \hline
     100 & {0 \\ 1 \\  2 \\3} & { $44.7725 - 0.271032$i \\ $44.7703 - 0.81311$i \\ $44.7658 - 1.35523$i \\ $44.7591 - 1.89743$i } & { $44.7725-0.271032$i \\ $44.7703-0.81311$i \\ $44.7658-1.35523$i \\ $44.7591-1.89743$i} & {$44.7726 - 0.222769$i \\ $44.7714 - 0.668316$i \\ $44.7692 - 1.11389$i \\ $44.7659 - 1.5595$i} & { \text{no results available} }\\
    \hline
    \end{tblr}
    \caption{Spin $s=2$ test-field QNM frequencies \textit{vs} physical gravitational QNM frequencies for $r_-=0.5$, computed  using the third order WKB approximation and Leaver's method.}
    \label{s2WKBcomp}
\end{table}

Second, let us look at spin $s=2$ test-field QNMs along with physical gravitational QNMs. We display results from \cite{Roussille:2023sdr} for the physical $l=2$ QNMs computed thanks the matrix continued fraction and we also ensured that our WKB results for the effective potential $V_{\text{phys}}$ with $l=2$ match with those displayed in \cite{Roussille:2023sdr}. As a complement we computed the physical QNM frequencies using the WKB method applied to $V_{\text{phys}}$ for $l=10$ and $l=100$. \\
Comparing the spin $s=2$ test-field QNMs computed with WKB on one side and the continued fraction on the other side, we observe that, as for the spin $s=0$, the accuracy gets better and better as $l$ becomes larger. 
One more comment can be made on the comparison with the physical QNMs computed using the WKB method: the real part of the physical QNMs and of the spin $s=2$ test-field QNMs becomes closer and closer when $l$ is large. This might mean that the additional physical terms in $V_{\text{phys}}$ do not influence the real part of QNMs in the eikonal limit $l\rightarrow \infty$.

\bibliographystyle{bib-style}
\bibliography{biblio.bib}
\end{document}